# s-Processing in the Galactic Disk.
# I. Super-Solar Abundances of Y, Zr, La, Ce in Young Open Clusters

E. Maiorca[1,2], S. Randich[2], M. Busso[1], L. Magrini[2], and S. Palmerini[1]

[1]Dipartimento di Fisica, Università di Perugia, and INFN, Sezione di Perugia; Via A. Pascoli, 06123 Perugia, Italy; emaiorca@gmail.com

[2]INAF, Osservatorio Astrofisico di Arcetri, Largo E. Fermi 5, 50125 Firenze, Italy

## ABSTRACT

In a recent study, based on homogeneous barium abundance measurements in open clusters, a trend of increasing [Ba/Fe] ratios for decreasing cluster age was reported. We present here further abundance determinations, relative to four other elements having important *s*-process contributions, with the aim of investigating whether the growth found for [Ba/Fe] is or not indicative of a general property, shared also by the other heavy elements formed by slow neutron captures. In particular, we derived abundances for yttrium, zirconium, lanthanum and cerium, using equivalent widths measurements and the MOOG code. Our sample includes 19 open clusters of different ages, for which the spectra were obtained at the ESO VLT telescope, using the UVES spectrometer. The growth previously suggested for Ba is confirmed for all the elements analyzed in our study. This fact implies significant changes in our views of the Galactic chemical evolution for elements beyond iron. Our results necessarily require that very low-mass AGB stars ($M \lessapprox 1.5 M_{\odot}$) produce larger amounts of *s*-process elements (hence activate the $^{13}$C-neutron source more effectively) than previously expected. Their role in producing neutron-rich elements in the Galactic disk has been so far underestimated and their evolution and neutron-capture nucleosynthesis should now be reconsidered.

## 1. Introduction

This paper presents photospheric abundance determinations of four elements synthesized primarily by slow neutron captures in a large sample of Galactic open clusters (OCs). Specifically, we measured the abundance of yttrium and zirconium, which belong to the first peak of the heavy-element abundance distribution beyond iron, near the neutron-magic number $N = 50$ (often called *light s-nuclei*, or *ls*) along with the abundance of lanthanum and cerium, which are close to the second peak, at $N = 82$ (*s*-process isotopes in this region are often called *heavy s-nuclei*, or *hs*). **Actually, the elements we consider here (not only certain isotopes in them) are usually called *s*-elements. This is certainly conditioned by what we know of the solar composition and does not necessarily apply to the Galaxy as a whole. As an example, it is known that the Ba abundance observed in the Galactic halo is primarily synthesized by the *r*-process. However, as we deal with the Galactic disk, and as our results will suggest that in Open Clusters the *s*-process component is even more important than in the Sun, we shall consider the usual definition of *s*-elements as appropriate to Y, Zr, La and Ce for our purposes.**

The clusters in our sample cover a wide range of ages and Galactocentric distances, which facts allow us to reconstruct the evolution of *s*-element abundances from the formation of the Galactic disk up to rather recent epochs. This should be compared with the predictions of Galactic chemical evolution models, which in turn require yields from nucleosynthesis computations in stellar models. This topic is therefore closely linked to the modeling of low mass stars (LMS), as they are the main contributors to the production of *s*-elements above $A \simeq 85$ (the so-called *main* component of the *s*-process) during their thermally-pulsing asymptotic giant branch (AGB) phases (see e.g. Busso et al. 1999).

Recently, D'Orazi et al. (2009) measured barium abundance in a large sample of OCs, finding for the first time the surprising result that barium increases with respect to iron for decreasing cluster age. This is at odds with expectations of previous Galactic chemical evolution models based on the $s$-process scenario consolidated in the nineties (see e.g. Raiteri et al. 1999; Travaglio et al. 1999). The results by D'Orazi et al. (2009) were based on relatively unevolved stars, which have inherited Ba from the previous evolution of the Galaxy and not produced it themselves. They showed that chemical evolution models could reproduce the observed trend only by introducing barium yields from very low mass stars enhanced with respect to estimates present in the literature (see Busso et al. 2001), despite these last had been previously shown to be adequate to reproduce the observed abundance trend up to the solar formation (Travaglio et al. 1999). In fact, D'Orazi et al. (2009) ascertained that they needed a larger production of Ba only from stars with masses between 1 and 1.5 $M_{\odot}$, which did not contribute significantly in determining the solar composition. Similar results were subsequently obtained by Jacobson et al. (2011a).

We refer to Busso et al. (1999) for a detailed description of the $s$-element nucleosynthesis in stars. Recent updates are in Straniero et al. (2009) and Cristallo et al. (2009). We briefly recall here that LMS, during AGB phases, produce $s$-elements mainly through the activation of the $^{13}C(\alpha,n)^{16}O$ reaction (the so-called $^{13}C$ neutron source). Producing $^{13}C$ in the He intershell during the interpulse periods requires the injection of protons from the envelope into layers below those fully mixed by the third dredge-up, through some form of extra-mixing. Afterwards, the previously formed $^{12}C$ nuclei can undergo proton captures creating $^{13}C$ seeds. Heavy nuclei near the valley of $\beta$-stability present in the He intershell, starting from the same Fe isotopes, can thus capture neutrons, activating $s$-processing. Gallino et al. (1998) showed that the efficiency of the $s$-process depends on metallicity, while in those years no relationship with the stellar mass (in the range 1.5 to 3 $M_{\odot}$) could be clearly ascertained. The results by D'Orazi et al. (2009) for Ba now seem to integrate this general view, suggesting that the efficiency of the $s$-process is larger in smaller masses ($M \leq 1.5 M_{\odot}$).

If this anti-correlation between the stellar mass and the effectiveness of neutron-capture nucleosynthesis exists, then one would expect the abundances of other $s$-elements to be affected in a way similar to Ba. In order to verify the above suggestions and to put solid constraints on *s*-process nucleosynthesis in the recent Galaxy, we derived the abundances of the four neutron capture elements mentioned above (Y, Zr, La, Ce) in a sample of stars from OCs, similar to the one analyzed by D'Orazi et al. (2009).

Our paper is structured as follows: in Sect. 2 we describe the OC sample, along with the methods of analysis used to derive our abundances. In Sect. 3 we present our main results. A preliminary discussion of the expected consequences is done in Sect. 4 and some conclusions are drawn in Sect. 5.

## 2. Sample and Analysis

### 2.1. Open Cluster Sample

EDITOR: PLACE TABLE 1 HERE.

The sample includes 19 OCs with ages in the range ∼ 0.1 – 8.4 Gyr and Galactocentric distances ($R_{GC}$) between 7.05 and 22 kpc. This sample has a big overlap with the one from D'Orazi et al. (2009); therefore we refer to this paper and to the references therein for additional information on the clusters in common and on their spectroscopy. In particular, we mention that we adopt here the same cluster ages used by D'Orazi et al. (2009); these are on a homogeneous scale and were taken from Magrini et al. (2009).

A few differences are however present in our sample; namely, we do not include the young clusters IC 2602, IC 2391, and NGC 6475, since the available spectral range did not cover useful lines for the abundance analysis. On the other hand, we added three new clusters: IC 4756, NGC 5822 (Pace et al. 2010), and NGC 6192 (Magrini et al. 2010). The spectra of the stars in these clusters were all collected with UVES, consistently with the rest of the sample. Details on the observations and instrument configuration can be found in Pace et al. (2010), and Magrini et al. (2010). Also, and most important, the adopted ages for the clusters we study are on the same scale as those of D'Orazi et al. (2009).

The full list of sample clusters is given in Table 1 along with their ages (Col. 2), Galactocentric distances (Col. 3), metallicities (Col. 4), numbers of stars analyzed, and spectral regions considered. We analyzed spectra of main sequence (MS) stars for the first eight OCs in the table, and spectra of red giant branch (RGB) or clump stars for the remaining ones, where MS stars are too faint to be observed with UVES.

### 2.2. Selected lines

EDITOR: PLACE TABLE 2 HERE.

In Table 2 we list the lines of Y, Zr, Ce and La used in the present analysis, along with the atomic parameters and their references. All the lines were very carefully inspected on the UVES spectrum of the Sun, and, in particular, the presence of possible blends was checked. Note that all lines belong to ionized species. Following previous works in the literature, we did not consider hyperfine structure (HFS) for Ce, Y, and Zr (Lawler et al. 2009; Hannaford et al. 1982; Sneden 1973). On the other hand, La needs HFS calculations and we adopted those used in Lawler et al. (2001). Finally, we mention that none of the lines is known to be affected by NLTE effects.

As anticipated above, for all the elements used in this analysis, we selected only the spectral features that can be detected in the UVES spectrum of the Sun, in order to be able to derive differential abundances line-by-line with respect to the Sun itself.

Equivalent widths (EWs) of the lines listed in Table 2 were measured using the *splot* routine within the IRAF package[1]. Whenever possible, measurements were performed by means of direct integration of lines instead of fitting with Gauss' or Voigt's profiles.

### 2.3. Solar abundances

EDITOR: PLACE TABLE 3 HERE.

Also for the Sun, abundances were derived from EW measurements using the MOOG code (Sneden 1973, we adopted the 2009 version) and the grid of one-dimensional local thermal equilibrium (LTE) model atmospheres by Kurucz (1993). Stark and radiative broadening were treated in the standard way and for collisional damping we adopted the classical Unsöld approximation. In Columns 4-5 of Table 2 we show our EW measurements for the Sun along with abundances in the usual form (A(X)=Log[N(X)/N(H)]+12). EWs were measured on the solar spectrum obtained with UVES. As in Randich et al. (2006), we adopted the following Solar parameters: $T_{\rm eff}$=5770 K, log $g$=4.44 and $\xi$ = 1.1 km/s.

We mention that, initially, we had also derived abundances using lines in the wavelength interval 300-390 nm, but the resulting values are systematically lower (~0.3dex) than those from lines with $\lambda$ >390 nm. Since the same behavior is obtained also by performing spectral synthesis (Sneden private communication), we believe that some problems in the opacity calculation arise in this UV region. Therefore we decided to discard these lines, limiting our analysis to those with wavelengths larger than 390 nm.

The mean solar abundances for the four elements derived from our analysis are shown in

---

[1]IRAF is distributed by the National Optical Astronomical Observatories, which are operated by the Association of Universities for Research in Astronomy, under contract with the National Science Foundation.

Table 3, together with the values derived by Asplund et al. (2009), hereafter AS09, using 3D models. The meteoritic abundances from Lodders et al. (2009) are also indicated.

Our abundances for Y, Zr, and Ce are in very good agreement with both the meteoritic values and AS09 determinations. On the other hand, the abundance of La derived in this study agrees with the value of AS09, but both are lower than in the meteoritic data. Although the difference is within the uncertainties (see Table 3), it is larger than the differences we found for other elements and indicates that additional studies on the solar lanthanum are required.

### 2.4. Abundances in OCs

We applied the same method used for the Sun to measure abundances in OC stars. Stellar parameters for the sample stars were taken from the reference papers given in Table 1, while in Table 6 we show their values. **We underline that all the stellar parameters of this study were determined spectroscopically**. As mentioned, our analysis of OC stars is strictly differential with respect to the Sun: thus, for each line we subtracted from the derived abundance the corresponding solar abundance from the same line, obtaining $[X/H]_{line}$. [X/H] values for each star were then obtained as the average of $[X/H]_{line}$ values. In this way, the uncertainties on the adopted set of atomic parameters, in particular $\log gf$ values, cancel out. The [X/Fe] estimate for each star was finally derived by subtracting the [Fe/H] value of the star itself (rather than the mean metallicity of the cluster) from its [X/H] value.

### 2.5. Errors

EDITOR: PLACE TABLE 4 HERE.

EDITOR: PLACE FIGURE 1 HERE.

Random errors include the contributions from errors on EW measurements and from uncertainties on stellar parameters.

Errors due to EW measurements in general correspond to the line-to-line abundance scatter for a given element, since the possible scatter due to uncertainties in $\log gf$ values is minimized by use of differential analysis with respect to the Sun. However, when less than three lines were available, we obtained five EW measurements and computed the corresponding average [El/H] values; the standard deviation of the measurements was then adopted as the error in [El/H].

As usually done, errors due to uncertainties in stellar parameters were estimated by varying each parameter separately, while leaving the others unchanged. The results are shown in Table 4; we performed this analysis for four stars that fairly well cover the parameter space of the whole sample. Uncertainties in effective temperature, surface gravity, and microturbulence were estimated following the discussions in Sestito et al. (2006) and in Randich et al. (2006). We found that variations of 100 K in $T_{eff}$ do not produce significant changes in the abundances, neither in red giants, nor in MS stars. On the other hand, and as is expected from an analysis based on lines from ionized elements, a 0.25 dex difference in $\log g$ leads to a variation of ±0.1 dex in the [X/H] value. Finally, changes of ±0.15 km/s in the micro-turbulence parameter $\xi$ of red giants have a rather large effect (about 0.15 dex) for Y, while they are less significant ($\sim$ 0.05 dex) for Ce (notice that we don't have measurements for Zr and La in red giants). The effect of a different $\xi$ is smaller for dwarf stars.

Systematic errors might be due to different effects such as the spectrocode, model atmospheres, or estimates of the continuum in the spectra, and of $\log gf$ values. While, as mentioned, the effect of the latter should be minimal thanks to the use of the differential analysis, the other three effects are more difficult to estimate. For three of the sample clusters measurements of one or more of the elements analyzed in this paper are available in the literature; namely M 67 (Tautvaisiene et al. 2000; Yong et al. 2005; Pancino et al. 2010; Önhag et al. 2010), Be 29 (Yong et al. 2005), the Hyades and Collinder 261 (De Silva et al. 2007), and IC 4651 (Mikolaitis et al. 2011). For one cluster only, M 67, more than one s-process determination is available in the literature. In Fig. 1 we compare our abundances for this cluster with three literature values. The figure indicates that our estimates generally agree within the errors and within ±0.05 dex with previous determinations; however, the two Zr measurements from the literature are considerably lower than our own value. We believe that the latter is likely more correct, since the significantly undersolar [Zr/Fe] derived by Tautvaisiene et al. (2000) and Yong et al. (2005) would not be consistent with the close-to-solar abundance ratios for Y, La, and Ce found by us and by other authors, and, more in general, with the solar composition of all the other elements. We finally mention that, in any case, possible systematic errors would not affect our results that are based on the comparison of clusters analyzed with the same method, and thus on the same abundance scale.

### *2.5.1. Gravity uncertainty*

EDITOR: PLACE TABLE 7 HERE.

**In this study we analyzed singly ionized lines for all the elements. Therefore gravity is important in our determinations, as can be seen from Table 4. In Table 5 one can notice a peculiarity. For some MS stars the** $\log\ g$ **values (that were derived from spectroscopic measurements available in the current literatures), are larger than expected on the basis of their temperature and evolutionary status. As typical examples, we quote KW 208 and KW 392 in the Praesepe cluster, with estimates for** $\log\ g$ **of 4.6 dex, and the two stars belonging to NGC 5822 with log g=** 4.7 **dex. For all of them we would rather** $\log\ g$ **values below 4.5, hence we decided to re-derive their parameters spectroscopically ourselves. We used the iron line list already adopted in Randich et al. (2006). EWs were measured with *daospec*, while abundances were obtained with MOOG (2009 version). Our results are**

**summarized in Table 7, where we listed also the number of Fe I and Fe II lines employed. They produced** $\log\ g$ **values smaller by** $0.2 - 0.3$ **dex than those of Pace et al. (2008) for KW 208 and KW 392. Using our estimates, the [El/H] abundances for these stars are reduced by less than** $\sim 0.1$ **dex, according to what was already shown in Table 4. However, the average abundances for Praesepe change only by** $\sim 0.05$ **dex with respect to those derived with the parameters measured by Pace et al. (2008). Concerning the two stars belonging to NGC 5822 we found a** $\log\ g$ **reduction of** $\sim 0.1$ **dex for TATM 11003 and of** $\sim 0.3$ **dex for TATM 11014. The subsequent cluster mean abundance is less than** $\sim 0.1$ **dex smaller than the value derived using stellar parameters from the literature. Therefore, the effects of these few peculiar cases is vanishingly small on our general trends, so that we can conclude that they do not affect in any significant way our conclusions.**

## 3. Results

EDITOR: PLACE TABLE 5 HERE.

Our results are summarized in Tables 5 and 6, where we list the [X/H] values for all the stars analyzed and the average [X/H] and [X/Fe] values for each of the target cluster, respectively. As discussed in Sect. 2.5, the errors listed in Table 6 refer to the line-to-line scatter; on the other hand, errors in Table 5 provide the standard deviation with respect to the cluster mean. We mention in passing that the mean values for Be 32 were computed excluding the star 941, whose abundances deviate by more than $2\sigma$ from the average. This star is discussed below in Sect. 4.2. Tables 5 and 6 put in evidence four major points which will be discussed in detail in the following sections. First, all the sample clusters show a rather homogeneous composition in the four *s*-process elements, although one star is present in Be 32 with abundances enriched with respect to the mean of the cluster which it belongs to. Second, clusters that appear enriched in first-peak elements (Y and Zr) are also enriched in second-peak elements (La and Ce). Third, clusters where dwarf members were analyzed show a similar enrichment in Y and Ce, while those where the abundances were based on the spectra of giants, together with Hyades stars (dwarfs stars), show higher enrichments in Y. Fourth, and most important, young clusters are generally characterized by higher *s*-element abundances than old clusters.

EDITOR: PLACE TABLE 6 HERE.

### 3.1. Ce vs. Y

EDITOR: PLACE FIGURE 2 HERE.

EDITOR: PLACE FIGURE 3 HERE.

In Fig. 2 we plot [Ce/H] vs [Y/H] for all the stars for which both elements were measured. Filled and open circles indicate dwarf and giant stars, respectively. The figure shows that the data points for dwarfs are clustered around the 1:1 relationship and are thus consistent with a similar average enrichment in Y and Ce, within uncertainties; on the other hand, the points of the giants have systematically lower Ce abundances than Y. Interestingly, the same holds also for the Hyades where dwarfs stars were been analyzed and where indeed Y appears to be more abundant than Ce. In these stars, the 508.743 nm yttrium line gives [Y/H] values consistent with [Ce/H] and systematically smaller than those from the other yttrium lines. The discrepancy is larger than 0.3 dex in some stars, characterized by low temperature (below 4500 K); namely, 2, 6, 7 in Cr 261 and 1240 in Be 20 and again the [Y/H] from the 508.743 nm is consistent with [Ce/H] (see Table 6). Moreover we noted that in all these stars the yttrium EWs are larger than 60 mÅ. In order to find some explanations to these observations, we considered the values [Y/Fe] derived from one yttrium line only (508.743 nm) for four OCs: M 67, Be 32, Mel 66 and NGC 6253. These clusters have ages between 4 and 6 Gyr. Fig. 3a shows our EWs measurements of the mentioned line versus the corresponding [Y/Fe] values. The points scatter around the center, at [Y/Fe]≈0. Performing the same exercise with the 520.041 nm yttrium line a very different behavior appears. We show it in Fig. 3b. The M 67 stars are centered on the [Y/Fe]≈0 value, and indeed the EWs are smaller than 60 mÅ. The other points have larger EWs and are considerably shifted to the right side of the plot, hence to higher abundances, forming a plateau. The same of the 520.041 nm happens for the remaining yttrium lines. The strict and systematic correlation between this shift to higher abundances and the EWs values for which it happens (EWs larger than about 60 mÅ), together with the shape of the trend of Fig. 3b are typical effects of saturation. Therefore in the following we shall use Y abundances from the 508.743 nm line only for the evolved stars and for the Hyades members, because this line does not suffer from saturation for EWs values we measured.

## 4. Discussion

### 4.1. Evolution with Age

EDITOR: PLACE FIGURE 4 HERE.

Fig. 4, where we plot [X/Fe] ratios vs. cluster age, summarizes our results for the evolution with age of the four elements analyzed in this study. The figure clearly shows that all the four elements, and specifically their logarithmic ratios to Fe, increase with decreasing cluster age. All [El/Fe] ratios are consistent with the solar value, or slightly below it, in old clusters. A steep growth is then seen between ages of $\sim$ 1.5 and 0.5 Gyr, after which the [El/Fe] values remain constant around a plateau value of +0.2 dex. The distribution and growth of the first-peak elements is very similar to that of the second-peak ones. Also, although lanthanum and zirconium have a smaller number of points than yttrium and cerium because of the available wavelength range of the star spectra of each cluster, we note that the measured [Zr/Fe] ratios are in very good agreement with [Y/Fe] in the clusters where both elements could be measured; the same happens for lanthanum and cerium. Our results confirm and strengthen those previously obtained by D'Orazi et al. (2009) for barium; indeed, our data are based on much higher statistics, as shown in Fig. 5. This, in turn, makes clear that there is a need for an *s*-element production from very low-mass AGB stars (with large ages, hence capable of affecting young clusters, not old ones) more effective than previously imagined.

EDITOR: PLACE FIGURE 5 HERE.

Together with the general trend of increasing *s*-element abundances in young clusters, some other features appear in the figure. First, we notice that at all ages a dispersion in *s*-process abundance is present. In particular, the Praesepe cluster lies significantly below the main trend and is less enriched than e.g., the coeval Hyades cluster. On the other hand, three clusters with ages around 0.8–1 Gyr, NGC 2660, NGC 5822, and NGC 3680 are more enriched than other members of the sample with similar ages. We have no secure explanation for this dispersion. We suggest that in part it might not be intrinsic. Rather, it may be due to wrong [Fe/H] determinations for these clusters. One example of this fact is probably shown by Praesepe, for which a high metallicity ([Fe/H]=0.27) was measured (Pace et al. 2008). Its [El/H] values are in general agreement with those of other coeval clusters, but due to this high [Fe/H] estimate, the [El/Fe] ratios become much smaller. This high [Fe/H] value of the Praesepe is actually rather strange for its Galactocentric radius, $R_{GC}$. This peculiarity makes Praesepe a very interesting target for new measurements of the metallicity. On the other hand, part of the dispersion in *s*-process elements might instead be intrinsic and be related to the different positions in the disk of the sample clusters (see the labeled OCs of Fig. 4 NGC 6192, IC 4651, NGC 6253, Cr 261 and the following paragraph).

We finally note the very high [Y/Fe] and [Ce/Fe] ratios of the old cluster Berkeley 29, both extremely above the trend. We will discuss the case of this cluster in a following section.

#### *4.1.1. [El/H] vs $R_{GC}$*

EDITOR: PLACE FIGURE 6 HERE.

EDITOR: PLACE FIGURE 7 HERE.

The *s*-element enhancement can also be put in evidence by looking at the trend of [El/H] with $R_{GC}$ for clusters in different age intervals. This is shown in Fig. 6 where the trends of [Y/H] and [Ce/H] with the Galactocentric radius are plotted. The figure suffers from low number statistics, in particular at large Galactocentric distances. Nevertheless it clearly shows that *s*-elements are characterized by a gradient along the Galactocentric radius similar to the one displayed by iron, iron-peak, and $\alpha$ elements (see Magrini et al. 2009). Most noticeably, two different gradients seem actually to be present. One is traced by older OCs (open circles of the figure), and one characterizes the younger objects (filled circles). The latter lays above the previous one, along the entire range of $R_{GC}$ (see Fig. 7), suggesting that the increase in the *s*-element enrichment occurs across the whole Galactic disk, once very low mass stars, with long evolutionary time scales, have started to contribute.

Moreover, looking at Fig. 7, the four labeled clusters with $R_{GC}$ smaller than $\sim$ 7.7 kpc, showing a different behavior with respect to the other OCs. These clusters, Cr 261, NGC 6253, IC 4651 and NGC 6192 are labeled also in Fig. 4, where the trend they trace is seen to remain slightly below the main one. A possible explanation could be related to the infall process occurred during the disk formation. Indeed, a rapid enrichment of iron in the inner zone of the disk, favored by mass infall from the halo, could shift the peak of star formation there towards higher [Fe/H] and [O/H]; the zones with larger $R_{GC}$ would be instead characterized by a SFR peaked at lower [Fe/H] values. Since a larger amount of iron and oxygen makes the occurrence of the third Dredge-up in AGB stars more difficult and reduces the neutron exposure, (see Straniero et al. 2003, and references therein), it is reasonable that in the inner zone of the disk the *s*-element production was more limited. Further studies are required to extend the statistics of *s*-element measurements in the inner OCs ($R_{GC} < 7.5$ kpc), while chemical evolutionary calculations at different $R_{GC}$ could verify the reliability of the hypothesis outlined above.

### *4.1.2. Berkeley 29*

EDITOR: PLACE FIGURE 8 HERE.

As mentioned above, both yttrium and cerium in this cluster show an anomalous enrichment. This result is not in agreement with the trend shown by other clusters, as it is evident in Fig. 4. Indeed, the age (4.3 Gyr) of this cluster should imply a solar [El/Fe] value. We also note that members of this cluster are characterized by a dispersion in both yttrium and cerium, ranging from moderate to extreme enrichment.

In order to further check the high abundances by means of a differential comparison, we superimposed the spectra of two stars with similar parameters: namely, star 933 of Be 29 and star 938 of Be 32.

Fig. 8 shows an example of this comparison for the yttrium line $\lambda = 508.742$nm, and supports the anomalous *s*-element enrichment found in Be 29. The $S/N$ ratios of the two spectra (40-50 for 933, 50-60 for 938) do not allow us to be confident on the reliability of our EWs measurements. Iron and $\alpha$-element abundances of this cluster (see Sestito et al. 2008) have values that are compatible with the solar ones, so that our *s*-element abundances look strange. On the other hand, in favor of the enrichment there is the fact that all the analyzed lines of the four stars in Be 29 indicate an overproduction of yttrium and cerium. We note that Be 29 is also enriched in Ba (see Fig. 5). Further investigations on *s*-element abundances in outer OCs are required and could become relevant clues for the knowledge of these external zones of the disk and their evolution.

#### *4.1.3. Clusters studied in the literature*

EDITOR: PLACE FIGURE 9 HERE.

Several recent studies focusing on abundance determinations in open clusters include the measurement of one or more s-process elements. In Fig. 9 we show the evolution of Y and Zr (upper panel) and Ce and La (bottom panel) obtained by using the most recent abundance determinations. Note that cluster ages are (or have been put by us) on the same scale as used for the open clusters discussed in this paper.

Measurements of *s*-process elements from the literature are highly non-homogeneous and difficult to compare with the present study, because different lines, stars, and methods have been used. Nevertheless, the figure shows an overall fair agreement, as data from the literature follow the same trend displayed by our own measurements. The agreement is particularly good for Y, La, and Ce, while Zr is more scattered; specifically, a couple of relatively young clusters (NGC 1883 and NGC 7789) have a low [Zr/Fe] ratio, while a few very old ones (NGC 7142, NGC 1193, NGC 188, Be 31) are characterized by enhanced ratios. We note that for NGC 7789 enhanced [Y/Fe], [La/Fe], and [Ce/Fe] have been reported, so that we believe that Zr was underestimated by Tautvaisiene et al. (2005). On the other hand, Zr determinations for NGC 1193, NGC 188, NGC 7142, and Be 31 come from Friel et al. (2010) and from Jacobson et al. (2008). In both papers the abundances of Zr were based on at most two Zr I lines; large variations are seen within each cluster and, as stressed by Friel et al. (2010) abundances are highly uncertain. While new measurements of Zr and of the other s-process elements for all these clusters are needed, we conclude that the data from the literature generally confirm the main results of our paper.

### 4.2. Anomalously enriched star

EDITOR: PLACE FIGURE 10 HERE.

Table 6 shows the presence of one anomalous star, which is significantly more *s*-element enriched than the other members of the same cluster. This star is 941 in Be 32. The discrepancy between its [El/H] values with respect to the mean of the cluster it belongs to is evident. This enhancement is present both for Y and Ce, thus indicating that it is most likely real and not due to problems with the analysis.

In order to further investigate this point, we superimposed the spectra of this anomalous star to that of another star with similar stellar parameters and belonging to the same cluster, namely the star 19; this allows us to check if the discrepancy occurs because of possible errors in the determination of parameters like the star gravity or temperature. In Fig. 10 we show an example of this exercise. The temperatures of the two objects are equal, while their log $g$ differ by 0.1 dex and their metallicities differ by 0.06 dex. The two spectra are perfectly in agreement with each other, with the exception of the Y II line $\lambda = 508.7$ nm, which is stronger in the 941 spectrum. The agreement in the other regions of the spectra indicates that errors in stellar parameter determinations cannot be responsible of the anomalies in the *s*-element abundances found in 941. We performed this test for the other cerium lines, finding the same results as for yttrium. We thus conclude that the enrichment of this star is real.

The above anomaly might be understood attributing the *s*-element enhancement to a mass transfer phenomenon in a binary system. If the anomalous star has a white dwarf companion, we might suppose that this last was once the primary component of the system, with an intermediate mass (e.g. 3−4 $M_\odot$ or more) and that it underwent *s*-processing during its AGB phase. Such a star would not have activated efficiently the $^{13}$C neutron source, but it would have nevertheless produced neutrons through the $^{22}$Ne($\alpha$,n)$^{25}$Mg reaction in the thermal pulses. This neutron source is actually the dominant one for all intermediate-mass (Lugaro et al. 2007) and high-mass stars (Busso & Gallino 1985; Raiteri et al. 1992). In this scenario, the secondary component we see today would be the equivalent of a Ba-star, i.e. would have accreted *s*-process rich material from the AGB primary.

## 5. Conclusions

- Our analysis shows that the growth of barium in young OCs, originally found by D'Orazi et al. (2009), is accompanied by a similar increase of other neutron capture elements belonging both to both the first and second *s*-process peaks (yttrium, zirconium, lanthanum and cerium). Specifically, we find that clusters older than about 1.5 Gyr have roughly solar [El/Fe] values, while younger clusters show an enhancement of about 0.2 dex. This growth terminates with a plateau for the youngest members of our sample, after which no further enrichment is seen. This suggests that the further increase in barium found by D'Orazi et al. (2009) should be attributed to other causes, affecting barium but not the abundance determinations of the other elements we have studied. NLTE effects, as mentioned by D'Orazi et al. (2009), could explain this behavior; a saturation of the barium lines they used is also a possibility.

- The confirmation of the *s*-element enrichment in young OCs needs now a theoretical basis, including an enhanced production of neutron-capture nuclei by very low mass stars ($M \leq 1.5 M_\odot$), so far often neglected in dealing with *s*-processing in the Galactic disk. Very recently, models of extra-mixing phenomena that might be promising for the formation of an extended $^{13}$C pocket in very low mass stars have been discussed (Busso et al. 2007;

Nordhaus et al. 2008). They are based on dynamo-driven magnetic buoyancy episodes in magnetically-active stars (like those discussed in Andrews et al. 1988). These phenomena should occur also on the AGB (Busso et al. 2010; Palmerini et al. 2011) and would not require special conditions other than the presence of magnetized structures lighter than the environment. In particular, molecular weight inversions induced by nuclear burning are not necessary, contrary to what is the case with the commonly-assumed mixing phenomena induced by thermohaline diffusion. Hence, they might operate also below the dredge-up in the interpulse phases of AGB stars, where nuclear sources are not present. Models of such mixing events and of the ensuing nucleosynthesis in LM-AGB stars should now be coupled to Galactic evolution models in order to ascertain whether the whole set of enhanced *s*-process abundances so far measured can be accounted for. We have already started new calculations with this specific scope.

- We underline the high relevance for our analysis of the rather detailed knowledge achieved for open clusters. We can derive their age through solid and homogeneous estimates. This was fundamental for obtaining our results. No clear trend would have been seen using only the usual [El/Fe] vs. [Fe/H] analysis.

- We identified one star that is anomalously *s*-process rich. We interpreted its anomaly as the consequence of a mass-transfer episode in a binary system, from a more massive, evolved companion which had already produced *s*-elements.

- Finally, we underline that the models with enhanced *s*-element production, now required to explain our data, should be verified on other test elements involved in the same He-shell burning phases where *s*-elements are produced. This is in particular the case of fluorine, whose abundance is enhanced in *s*-process rich AGB stars (Abia et al. 2010). Observations aimed at measuring the F abundance also in MS and RGB stars of open clusters should be carried out to allow for a more complete set of constraints to the upgrade of AGB nucleosynthesis models that is now clearly required.

We are grateful to C. Sneden and S. Andrievsky for their help during the abundance analysis. We also thank G. Pace for providing spectra of members of IC 4756, NGC 5822, Praesepe, M 67, NGC 3680 and IC 4651. We acknowledge clarifying discussions on the *s*-process and on AGB nucleosynthesis with R. Gallino, O. Straniero, S. Cristallo, and C. Abia.

This manuscript was prepared with the AAS LaTeX macros v5.2.

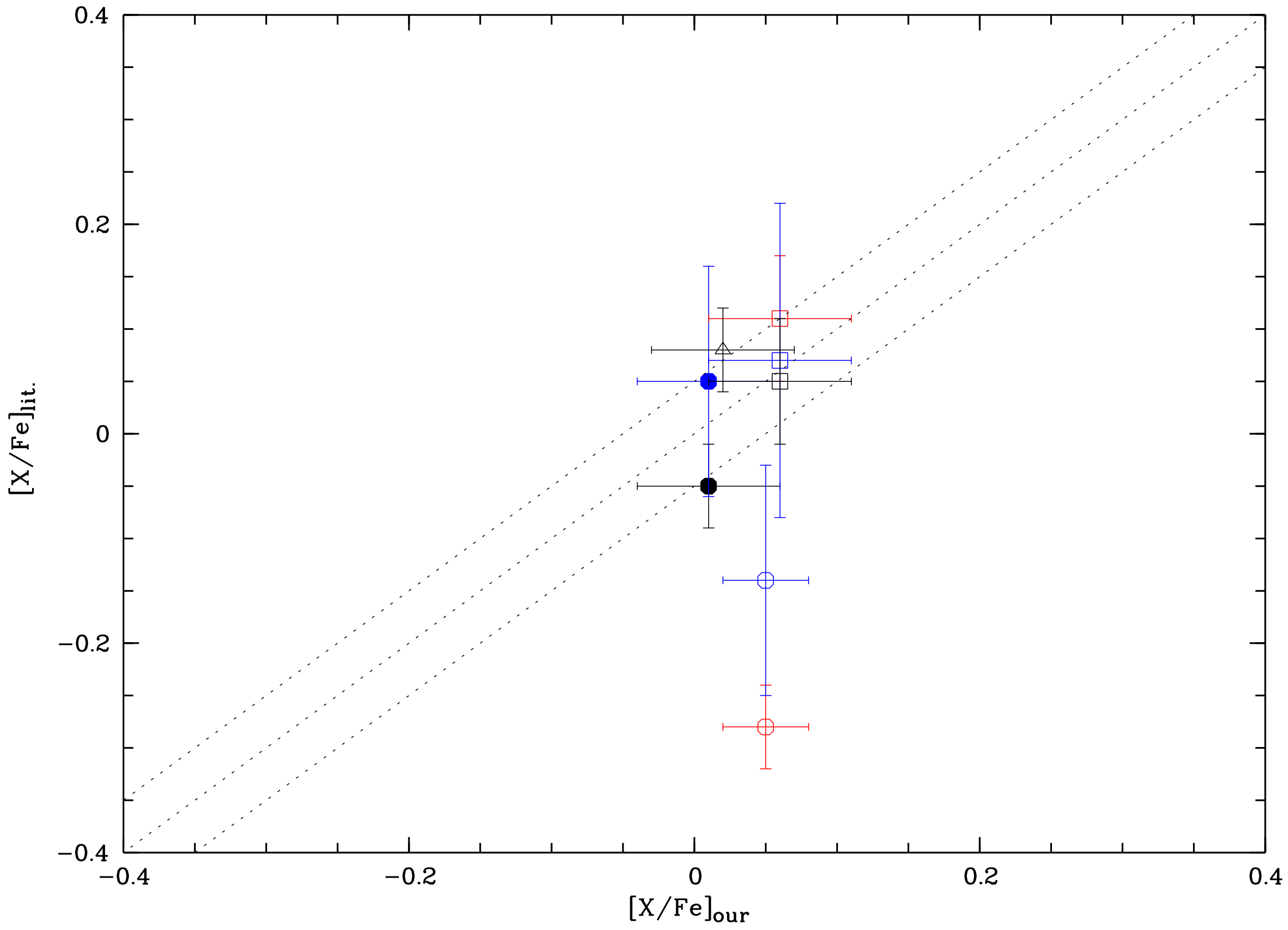

Fig. 1.— Comparison of our [X/Fe] determinations for M 67 with data available in the literature. Different symbols denote different elements, while different colors indicate different studies. Namely: filled circles, open circles, open triangles and open squares represent Y, Zr, La, and Ce, respectively; black, red, and blue denote Pancino et al. (2009), Yong et al. (2005) and Tautvaisiene et al. (2000). The dotted lines represent the $[Fe/H]_{our}$=$[Fe/H]_{lit.}$ ± 0.05 relation.

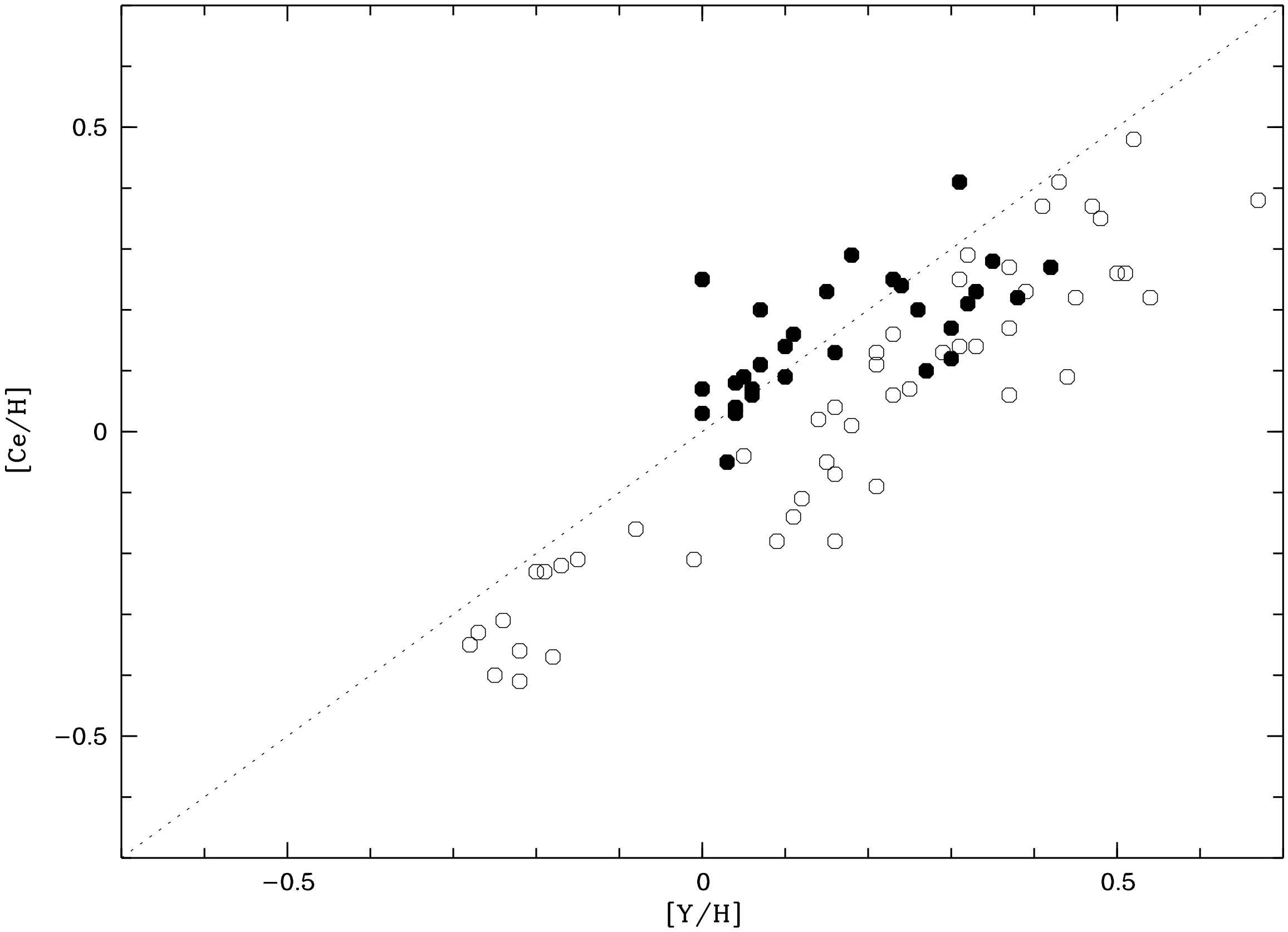

Fig. 2.— [Ce/H] vs. [Y/H]. Filled symbols denote dwarf stars, while open symbols indicate giant cluster members. The dashed line indicates the 1:1 relationship.

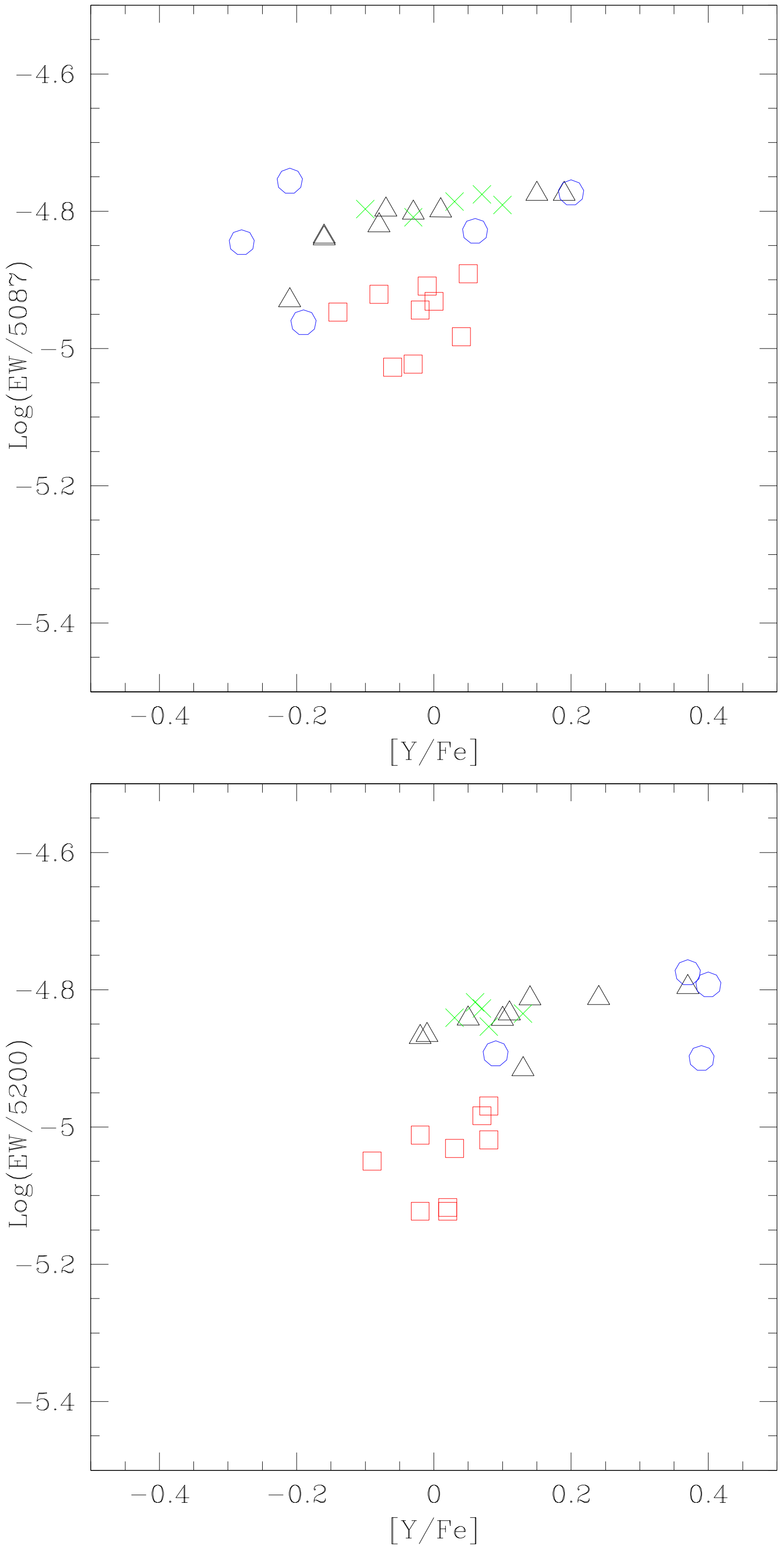

Fig. 3.— EWs measurements versus the corresponding abundances are plotted for two yttrium lines in four OCs: triangles refer to Be 32, squares to M 67, crosses to Mel 66 and circles to NGC 6253. a) Top panel: data from the 508.742 nm line. b) Bottom panel: data from the 520.041 nm line.

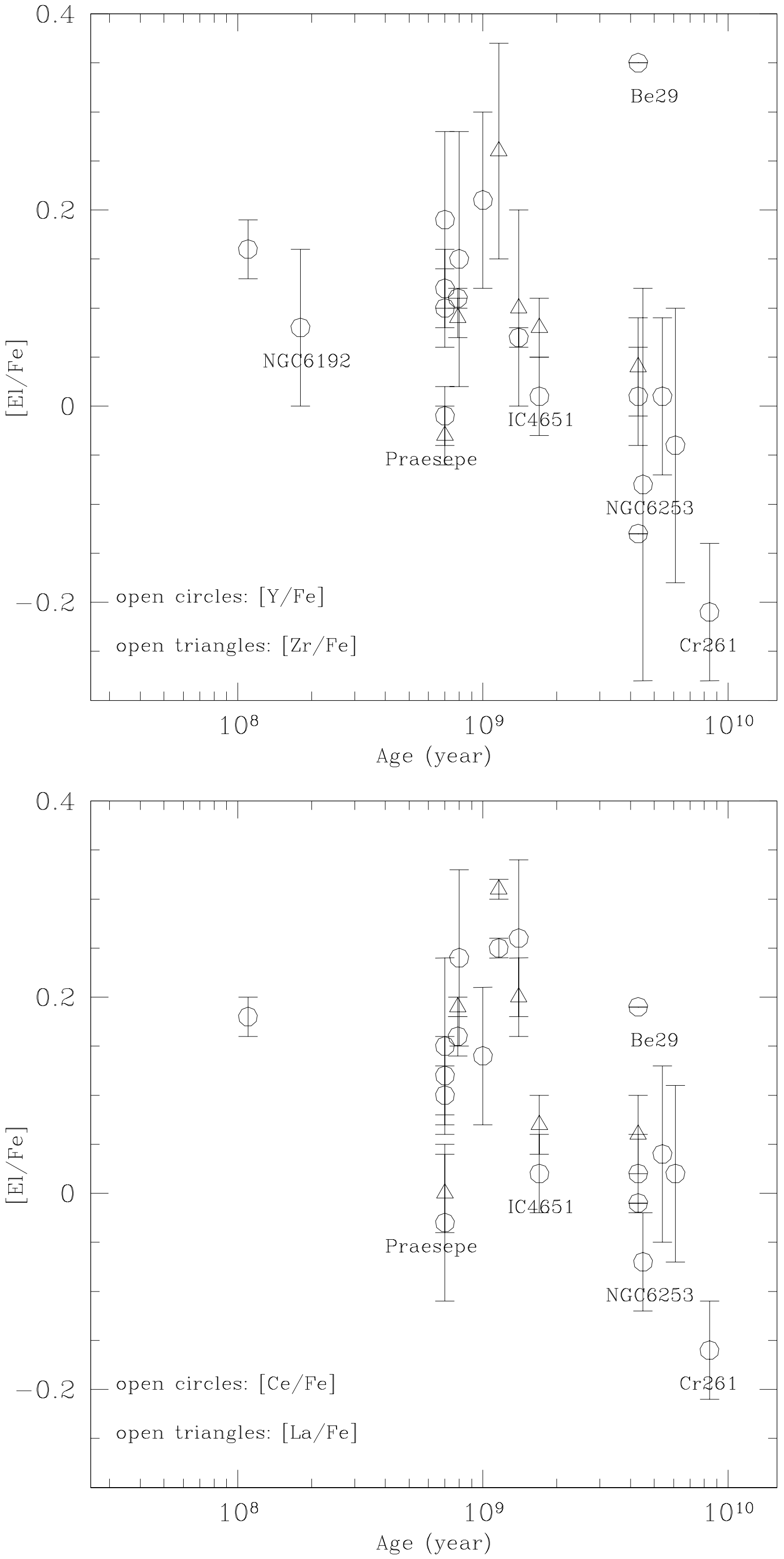

Fig. 4.— Average cluster [El/Fe] vs. cluster age. In the top panel we show the evolution of first peak elements, namely, [Y/Fe] (circles) and [Zr/Fe] (triangles); in the bottom panel the second peak elements La (triangles) and Ce (circles) are shown. Error bars for Be 20 and Be 29 are not illustrated for graphical reasons.

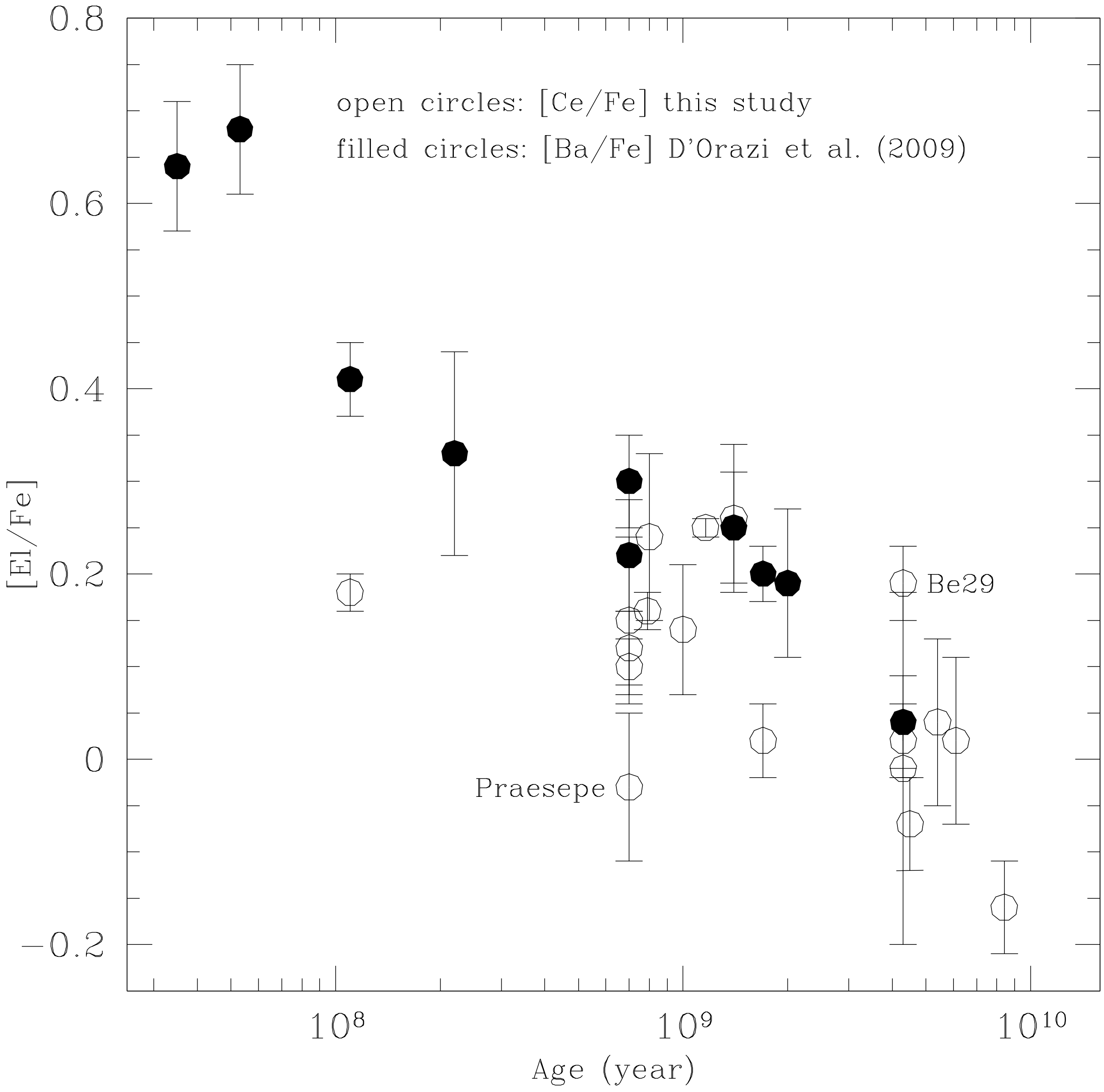

Fig. 5.— A comparison of the evolutionary trends with cluster age for [Ce/Fe] and [Ba/Fe].

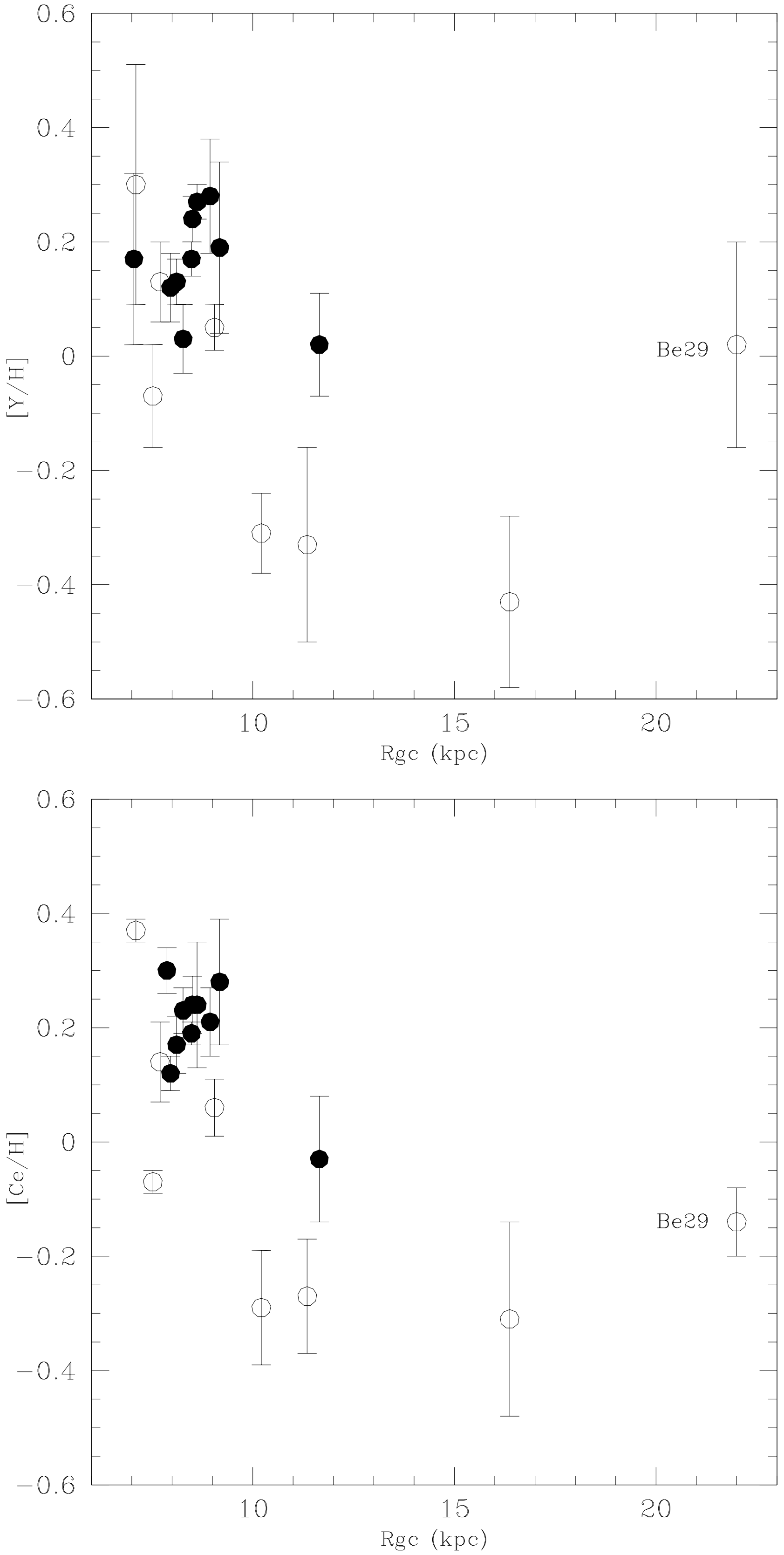

Fig. 6.— [El/H] versus Galactocentric radius for yttrium and cerium. Filled and open circles represent clusters younger and older than 1.5 Gyr, respectively.

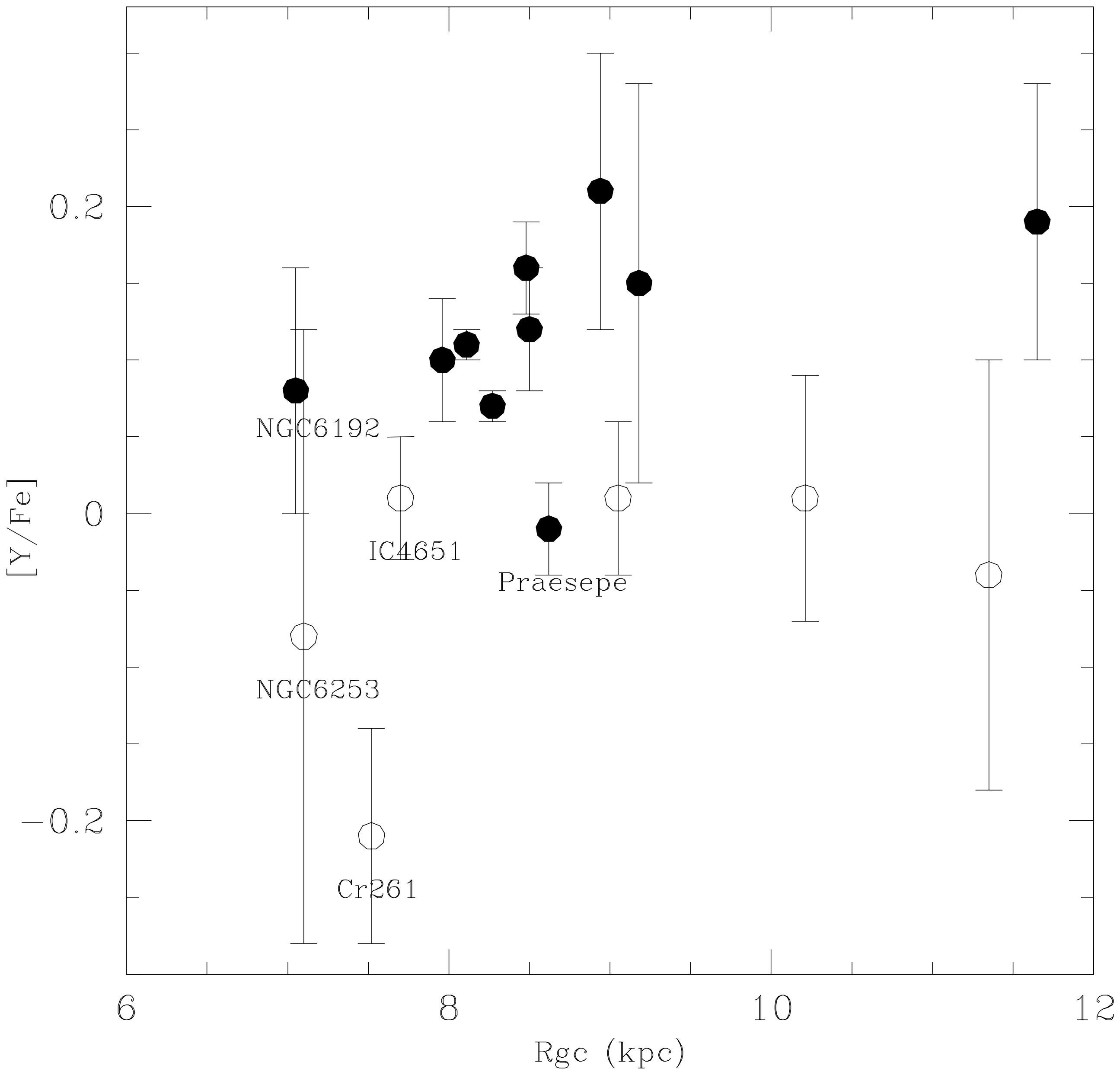

Fig. 7.— [Y/Fe] versus Galactocentric radius. Filled and open circles represent clusters younger and older than 1.5 Gyr, respectively.

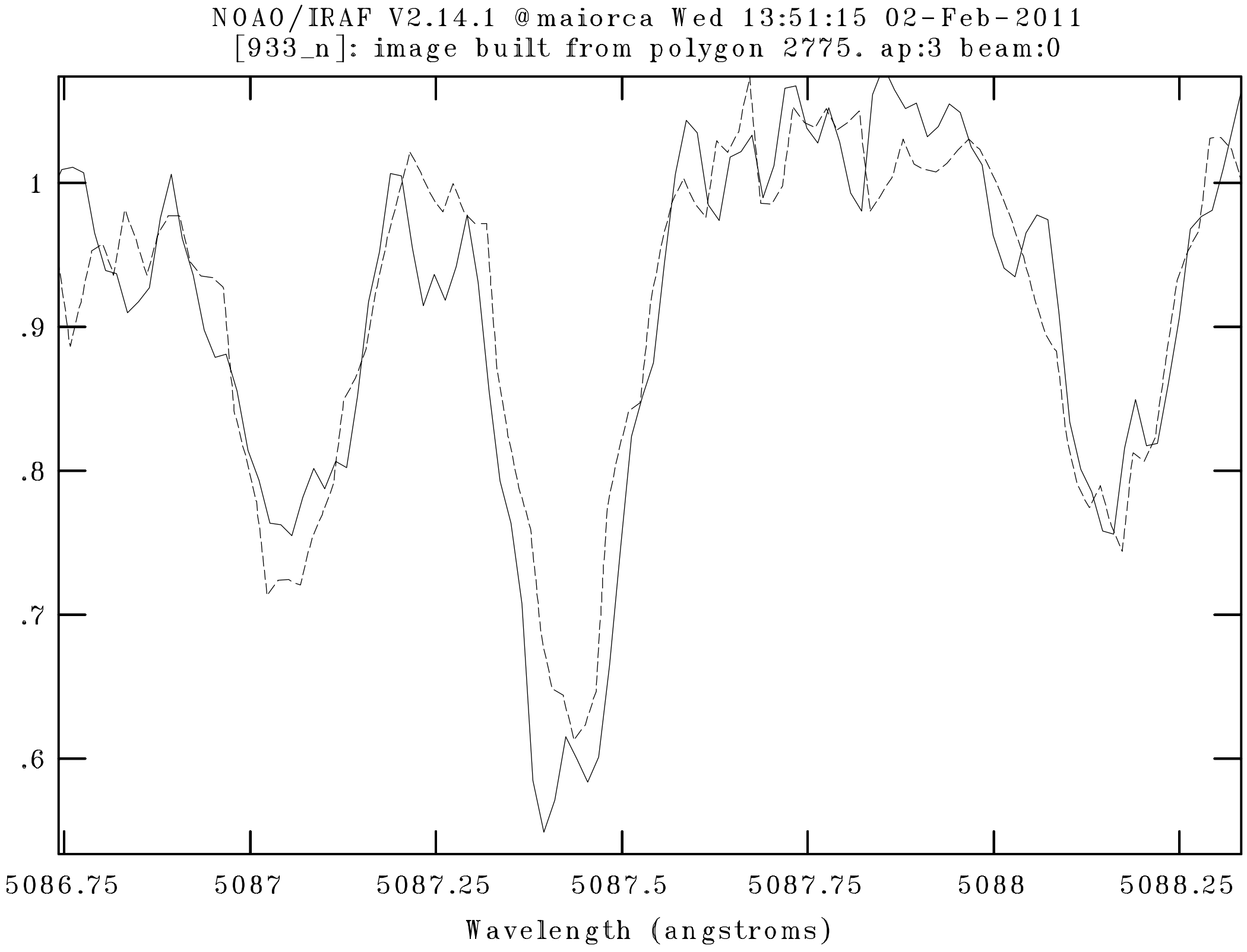

Fig. 8.— The yttrium line $\lambda = 508.743$ nm from the star 933 in Be 29 (solid line) and from the star 938 in Be 32 (dashed line). The comparison shows the larger EW of the line in the spectrum of 933.

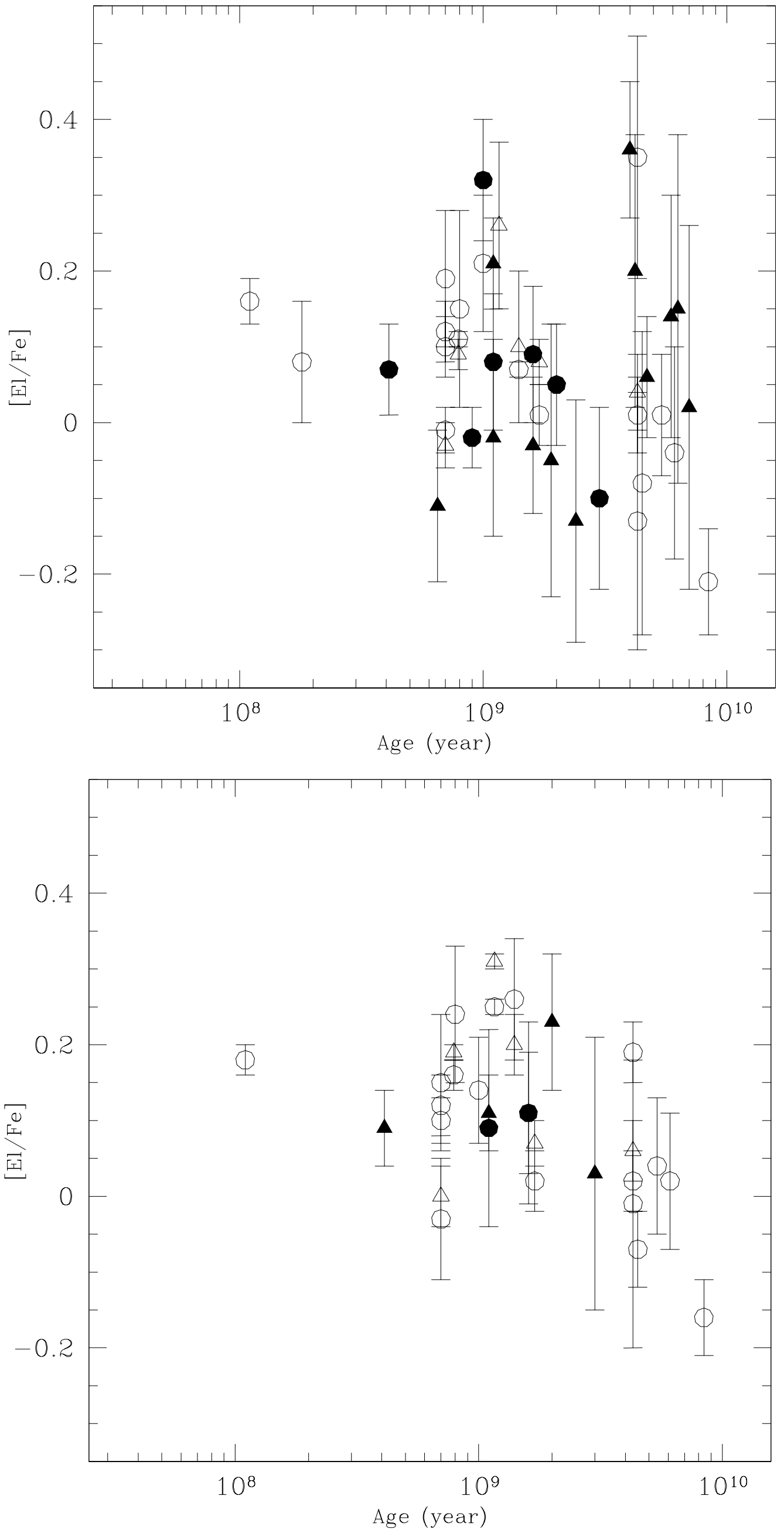

Fig. 9.— Similar to Fig. 4 but measurements from the literature are also shown. In the top panel we show the evolution of first peak elements, namely, [Y/Fe] (circles) and [Zr/Fe] (triangles); in the bottom panel the second peak elements La (triangles) and Ce (circles) are shown. Open and filled symbols denote abundances from this paper and the literature, respectively. Literature data include: Hamdani et al. (2000, NGC 2360); Tautvaisiene et al. (2005, NGC 7789); Carraro et al. (2008, NGC 2112); Jacobson et al. (2008, NGC 7142); Jacobson et al. (2009, NGC 1817, NGC 1883, NGC 2141, NGC 2158); Friel et al. (2010, Be 31, Be 39, NGC 1193, NGC 188); Pancino et al. (2010, NGC 7789, NGC 2420, Cr 110, NGC 2209); Mikolaitis et al. (2010, NGC 6134); Jacobson et al. (2011b, NGC2243)

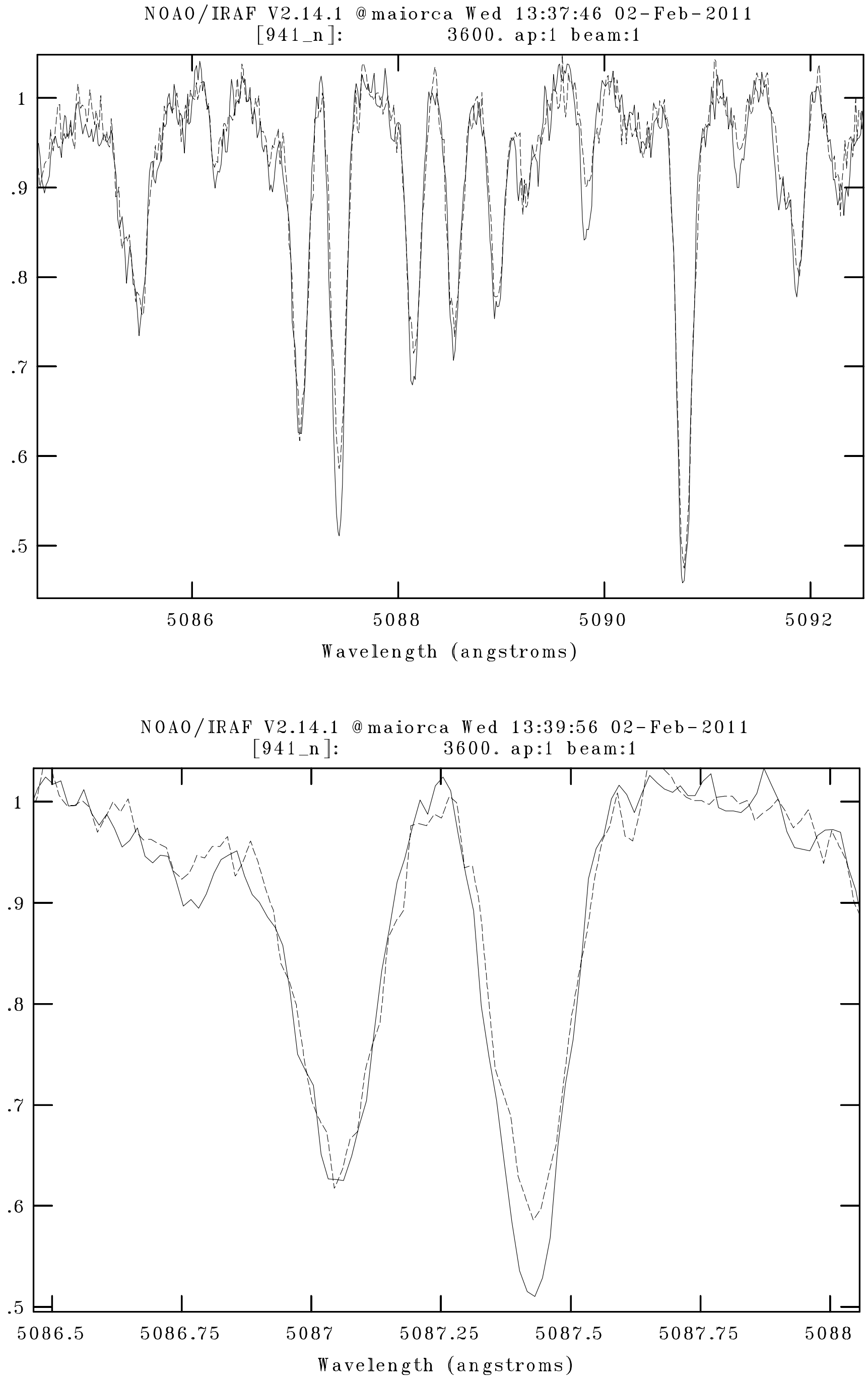

Fig. 10.— A comparison between the spectra of stars 941 and 19 in Be 32. The solid line represents star 941, the dashed one star 19. The Y II line is that at 508.743 nm. The upper plot shows the agreement between the two spectra, as a consequence of the similarity in the stellar parameters. The bottom panel is a zoom of the upper one and makes clear that there exists a discrepancy between the yttrium lines of the two stars.

Table 1. Our Open Clusters sample Sample

| Open Cluster | Age (Gyr) | $R_{GC}$ | [Fe/H] | References | # of stars | Spectral region (nm) |
|---|---|---|---|---|---|---|
| NGC 2516 | 0.11 | 8.48 | +0.01 ± 0.07 | 1,2 | 3 | 480–680 |
| Hyades | 0.7 | 8.50 | +0.13 ± 0.05 | 3,2 | 4 | 480–680 |
| Praesepe | 0.7 | 8.62 | +0.27 ± 0.04 | 4 | 7 | 330–680 |
| IC 4756 | 0.79 | 8.11 | +0.01 ± 0.04 | 5 | 3 | 330–450 |
| NGC 5822 | 1.16 | 7.87 | +0.05 ± 0.03 | 5 | 2 | 330–450 |
| NGC 3680 | 1.4 | 8.27 | −0.04 ± 0.03 | 4 | 2 | 330–680 |
| IC 4651 | 1.7 | 7.70 | +0.12 ± 0.05 | 4 | 5 | 330–680 |
| M67 | 4.3 | 9.05 | +0.03 ± 0.01 | 6,4 | 15 | 330–680 |
| NGC 6192 | 0.18 | 7.05 | +0.12 ± 0.04 | 7 | 4 | 480–680 |
| NGC 2324 | 0.7 | 11.65 | −0.17 ± 0.05 | 8 | 6 | 480–680 |
| NGC 3960 | 0.7 | 7.96 | +0.02 ± 0.04 | 9 | 5 | 480–680 |
| NGC 2660 | 0.8 | 9.18 | +0.04 ± 0.04 | 9 | 5 | 480–680 |
| NGC 2477 | 1.0 | 8.94 | +0.07 ± 0.03 | 8 | 6 | 480–680 |
| Berkeley 29 | 4.3 | 22.0 | −0.31 ± 0.03 | 10 | 4 | 480–680 |
| Berkeley 20 | 4.3 | 16.4 | −0.30 ± 0.02 | 10 | 2 | 480–680 |
| NGC 6253 | 4.5 | 7.10 | +0.39 ± 0.08 | 11 | 5 | 480–680 |
| Melotte 66 | 5.4 | 10.21 | −0.33 ± 0.03 | 10 | 6 | 480–680 |
| Berkeley 32 | 6.1 | 11.35 | −0.29 ± 0.04 | 9 | 9 | 480–680 |
| Collinder 261 | 8.4 | 7.52 | +0.13 ± 0.05 | 10 | 7 | 480–680 |

Note. — We list: cluster ages (Col.2); the Galactocentric distance (Col.3); this has been computed assuming a solar Galactocentric distance $R_{GC\odot}$ =8.5 kpc; [Fe/H] values (Col 4); references for the metallicity and stellar parameters (Col. 5); number of stars and wavelength range of the spectra analyzed (Cols. 6 and 7).

References. — 1. Terndrup et al. (2002); 2. Randich et al. (2007); 3. Paulson et al. (2003); 4. Pace et al. (2008); 5. Pace et al. (2010); 6. Randich et al. (2006); 7. Magrini et al. (2010); 8. Bragaglia et al. (2008); 9. Sestito et al. (2006); 10. Sestito et al. (2008); 11. Sestito et al. (2007)

Table 2. Line list and Solar abundances line-by-line

| Wavelength (nm) | EP (eV) | log $gf$ | $EW_{\odot}$ (mÅ) | $A(X)_{\odot}$ |
|---|---|---|---|---|
| | | Y II[a] | | |
| 412.491 | 0.41 | −1.50 | 22.8 | 2.24 |
| 439.801 | 0.13 | −1.00 | 51.5 | 2.18 |
| 485.487 | 0.99 | −0.38 | 48.1 | 2.26 |
| 488.369 | 1.08 | +0.07 | 59.1 | 2.18 |
| 490.012 | 1.03 | −0.09 | 56.6 | 2.23 |
| 508.743 | 1.08 | −0.17 | 49.4 | 2.15 |
| 520.041 | 0.99 | −0.57 | 38.2 | 2.18 |
| | | Zr II[b] | | |
| 405.032 | 0.71 | −1.00 | 23.2 | 2.53 |
| 420.898 | 0.71 | −0.46 | 44.0 | 2.52 |
| 437.974 | 1.53 | −0.25 | 25.5 | 2.60 |
| | | La II[c] | | |
| 398.857 | 0.40 | −0.455 | 51.3 | 1.12 |
| 399.577 | 0.17 | −0.686 | 39.0 | 1.08 |
| 408.671 | 0.00 | −0.696 | 40.5 | 1.12 |
| 433.377 | 0.17 | −0.686 | 48.1 | 1.20 |
| | | Ce II[d] | | |
| 399.924 | 0.29 | +0.060 | 18.8 | 1.56 |
| 404.258 | 0.49 | +0.000 | 10.6 | 1.49 |
| 407.347 | 0.48 | +0.210 | 18.8 | 1.58 |
| 408.322 | 0.70 | +0.270 | 18.0 | 1.71 |
| 413.765 | 0.52 | +0.400 | 29.8 | 1.74 |
| 414.500 | 0.70 | +0.100 | 10.0 | 1.54 |
| 434.979 | 0.70 | −0.320 | 5.4 | 1.64 |
| 518.746 | 1.21 | +0.170 | 5.3 | 1.58 |

Table 2—Continued

| Wavelength (nm) | EP (eV) | log $gf$ | $EW_{\odot}$ (mÅ) | $A(X)_{\odot}$ |
|---|---|---|---|---|
| 527.423 | 1.04 | +0.130 | 7.3 | 1.61 |

Note. — References for atomic parameters: (a) Hannaford et al. (1982); (b) Malcheva et al. (2006); (c) Lawler et al. (2001); (d) Lawler et al. (2009).

Table 3. Abundances in the Sun

| Element | Photosphere | AS09 | Meteorites |
|---|---|---|---|
| Y | 2.20 ± 0.04 | 2.21 ± 0.04 | 2.19 ± 0.04 |
| Zr | 2.55 ± 0.04 | 2.58 ± 0.04 | 2.55 ± 0.04 |
| La | 1.13 ± 0.05 | 1.10 ± 0.04 | 1.19 ± 0.02 |
| Ce | 1.61 ± 0.09 | 1.58 ± 0.04 | 1.60 ± 0.02 |

Table 4. Error estimates due to uncertainties in stellar parameters

| Δ[X/H] | $\Delta T_{\rm eff}$ = ±100 K | $\Delta \log g$ = ±0.25 *dex* | $\Delta\xi$ = ±0.15 $kms^{-1}$ | |
|---|---|---|---|---|
| Δ[Y/H] | 0.00/-0.01 | 0.08/-0.11 | -0.15/0.15 | (a) |
| Δ[Ce/H] | 0.01/0.00 | 0.11/-0.11 | -0.03/0.06 | (a) |
| Δ[Y/H] | 0.00/-0.02 | 0.08/-0.11 | -0.16/0.15 | (b) |
| Δ[Ce/H] | 0.01/-0.02 | 0.10/-0.11 | -0.05/0.06 | (b) |
| Δ[Y/H] | 0.02/-0.02 | 0.10/-0.10 | -0.06/0.07 | (c) |
| Δ[Ce/H] | 0.02/-0.02 | 0.10/-0.10 | 0.00/0.01 | (c) |
| Δ[La/H] | 0.03/-0.02 | 0.11/-0.09 | -0.01/0.02 | (d) |
| Δ[Zr/H] | 0.02/-0.02 | 0.10/-0.10 | -0.02/0.03 | (d) |

Note. — (a) Star 18 of Be32: $T_{\rm eff}$ = 4850 K, logg=2.3, $\xi$ = 1.21 $kms^{-1}$; (b) Star 10488 of NGC 2324: $T_{\rm eff}$ = 5100 K, logg=2.2, $\xi$ = 1.21 $kms^{-1}$; (c) Star S994 of M67: $T_{\rm eff}$ = 6151 K, logg=4.1, $\xi$ = 1.45 $kms^{-1}$; (d) Star sand1092 of M67: $T_{\rm eff}$ = 6160 K, logg=4.41, $\xi$ = 1.42 $kms^{-1}$

Table 5. Mean abundances in OCs

| Cluster | [Y/H] | [Y/Fe] | [Zr/H] | [Zr/Fe] | [La/H] | [La/Fe] | [Ce/H] | [Ce/Fe] |
|---|---|---|---|---|---|---|---|---|
| NGC 2516 | 0.17 ± 0.03 | 0.16 ± 0.03 | — | — | — | — | 0.19 ± 0.02 | 0.18 ± 0.02 |
| Hyades | 0.24 ± 0.04 | 0.12 ± 0.04 | — | — | — | — | 0.24 ± 0.05 | 0.12 ± 0.04 |
| Praesepe | 0.27 ± 0.03 | −0.01 ± 0.03 | 0.24 ± 0.05 | −0.03 ± 0.03 | 0.28 ± 0.06 | 0.00 ± 0.04 | 0.24 ± 0.11 | −0.03 ± 0.08 |
| IC 4756 | 0.13 ± 0.04 | 0.11 ± 0.01 | 0.10 ± 0.02 | 0.09 ± 0.02 | 0.20 ± 0.04 | 0.19 ± 0.01 | 0.17 ± 0.05 | 0.16 ± 0.02 |
| NGC 5822 | — | — | 0.30 ± 0.07 | 0.26 ± 0.11 | 0.36 ± 0.04 | 0.31 ± 0.01 | 0.30 ± 0.04 | 0.25 ± 0.01 |
| NGC 3680 | 0.03 ± 0.06 | 0.07 ± 0.01 | 0.07 ± 0.05 | 0.10 ± 0.10 | 0.17 ± 0.01 | 0.20 ± 0.04 | 0.23 ± 0.04 | 0.26 ± 0.08 |
| IC 4651 | 0.13 ± 0.07 | 0.01 ± 0.04 | 0.20 ± 0.07 | 0.08 ± 0.03 | 0.19 ± 0.04 | 0.07 ± 0.03 | 0.14 ± 0.07 | 0.02 ± 0.04 |
| M67 | 0.05 ± 0.04 | 0.01 ± 0.05 | 0.06 ± 0.05 | 0.04 ± 0.05 | 0.08 ± 0.04 | 0.06 ± 0.04 | 0.06 ± 0.05 | 0.02 ± 0.04 |
| NGC 6192 | 0.17 ± 0.15 | 0.08 ± 0.08 | — | — | — | — | — | — |
| NGC 2324 | 0.02 ± 0.09 | 0.19 ± 0.09 | — | — | — | — | −0.03 ± 0.11 | 0.15 ± 0.09 |
| NGC 3960 | 0.12 ± 0.06 | 0.10 ± 0.04 | — | — | — | — | 0.12 ± 0.03 | 0.10 ± 0.03 |
| NGC 2660 | 0.19 ± 0.15 | 0.15 ± 0.13 | — | — | — | — | 0.28 ± 0.11 | 0.24 ± 0.09 |
| NGC 2477 | 0.28 ± 0.10 | 0.21 ± 0.09 | — | — | — | — | 0.21 ± 0.06 | 0.14 ± 0.07 |
| Berkeley 29 | 0.02 ± 0.18 | 0.35 ± 0.16 | — | — | — | — | −0.14 ± 0.06 | 0.19 ± 0.04 |
| Berkeley 20 | −0.43 ± 0.15 | −0.13 ± 0.17 | — | — | — | — | −0.31 ± 0.17 | −0.01 ± 0.19 |
| NGC 6253 | 0.30 ± 0.21 | −0.08 ± 0.20 | — | — | — | — | 0.37 ± 0.02 | −0.07 ± 0.05 |
| Melotte 66 | −0.31 ± 0.07 | 0.01 ± 0.08 | — | — | — | — | −0.29 ± 0.10 | 0.04 ± 0.09 |
| Berkeley 32 | −0.33 ± 0.17 | −0.04 ± 0.14 | — | — | — | — | −0.27 ± 0.10 | 0.02 ± 0.09 |
| Collinder 261 | −0.07 ± 0.09 | −0.21 ± 0.07 | — | — | — | — | −0.07 ± 0.02 | −0.16 ± 0.05 |

Table 6. Differential abundances star by star. Values in bracket refer to the number of used lines to get the corresponding abundance

| Cluster | Star | $T_{\rm eff}$ | log $g$ | $\xi$ | [Fe/H] | [Y/H]$_{5087}$ | [Zr/H] | [La/H] | [Ce/H] | [Y/H]$_{\rm all}$ |
|---|---|---|---|---|---|---|---|---|---|---|
| NGC 2516 | CTIO2 | 5238 | 4.5 | 0.89 | 0.01 | +0.14 ± 0.03 (1) | — | — | +0.21 ± 0.04 (1) | +0.18 ± 0.04 (5) |
| | CTIO3 | 5110 | 4.5 | 0.85 | 0.01 | +0.16 ± 0.03 (1) | — | — | +0.20 ± 0.06 (2) | +0.33 ± 0.12 (5) |
| | DK213 | 5305 | 4.5 | 0.91 | 0.01 | +0.20 ± 0.04 (1) | — | — | +0.17 ± 0.03 (1) | +0.32 ± 0.10 (5) |
| HYADES | vb42 | 5308 | 4.5 | 0.91 | 0.10 | +0.24 ± 0.04 (1) | — | — | +0.17 ± 0.05 (2) | +0.30 ± 0.06 (5) |
| | vb109 | 5141 | 4.5 | 0.86 | 0.13 | +0.27 ± 0.04 (1) | — | — | +0.27 ± 0.04 (2) | +0.42 ± 0.09 (5) |
| | vb182 | 5079 | 4.5 | 0.84 | 0.13 | +0.19 ± 0.04 (1) | — | — | +0.28 ± 0.07 (2) | +0.35 ± 0.09 (5) |
| | vb187 | 5339 | 4.5 | 0.92 | 0.11 | +0.26 ± 0.04 (1) | — | — | +0.22 ± 0.07 (2) | +0.38 ± 0.08 (5) |
| Praesepe | KW49 | 6150 | 4.5 | 1.41 | 0.22 | +0.20 ± 0.04 (1) | +0.24 ± 0.11 (3) | +0.26 ± 0.18 (5) | +0.24 ± 0.07 (1) | +0.24 ± 0.14 (3) |
| | KW100 | 6150 | 4.3 | 1.78 | 0.27 | — | — | +0.20 ± 0.12 (2) | — | — |
| | KW208 | 6280 | 4.6 | 1.52 | 0.28 | +0.23 ± 0.03 (1) | +0.28 ± 0.10 (3) | +0.32 ± 0.15 (5) | — | +0.28 ± 0.05 (4) |
| | KW326 | 5800 | 4.5 | 1.28 | 0.29 | +0.18 ± 0.03 (1) | +0.22 ± 0.06 (3) | +0.28 ± 0.10 (5) | +0.25 ± 0.13 (6) | +0.23 ± 0.03 (5) |
| | KW368 | 5720 | 4.5 | 1.12 | 0.26 | +0.19 ± 0.05 (1) | +0.21 ± 0.06 (3) | +0.26 ± 0.06 (5) | +0.20 ± 0.10 (8) | +0.26 ± 0.05 (5) |
| | KW392 | 6250 | 4.6 | 1.48 | 0.35 | +0.31 ± 0.03 (1) | +0.31 ± 0.09 (3) | +0.37 ± 0.08 (5) | +0.41 ± 0.10 (6) | +0.31 ± 0.06 (5) |
| | KW418 | 6150 | 4.4 | 1.27 | 0.24 | +0.21 ± 0.04 (1) | +0.19 ± 0.08 (3) | — | +0.10 ± 0.09 (2) | +0.27 ± 0.07 (5) |
| IC 4756 | HER97 | 6118 | 4.5 | 1.29 | 0.00 | — | +0.10 ± 0.03 (3) | +0.20 ± 0.05 (5) | +0.14 ± 0.05 (3) | — |
| | HER165 | 6070 | 4.5 | 1.24 | 0.05 | — | +0.11 ± 0.13 (3) | +0.23 ± 0.01 (5) | +0.23 ± 0.05 (7) | +0.15 ± 0.07 (2) |
| | HER240 | 6007 | 4.5 | 1.21 | -0.02 | — | +0.08 ± 0.04 (3) | +0.16 ± 0.06 (5) | +0.14 ± 0.10 (7) | +0.10 ± 0.07 (2) |
| NGC 5822 | TATM11003 | 6160 | 4.7 | 1.05 | 0.02 | — | +0.35 ± 0.09 (3) | +0.33 ± 0.08 (5) | +0.27 ± 0.06 (4) | — |
| | TATM11014 | 6273 | 4.7 | 1.26 | 0.07 | — | +0.25 ± 0.04 (3) | +0.38 ± 0.03 (5) | +0.32 ± 0.04 (2) | — |
| NGC 3680 | AHTC1009 | 6010 | 4.5 | 1.16 | 0.00 | +0.09 ± 0.03 (1) | +0.03 ± 0.08 (3) | +0.17 ± 0.04 (5) | +0.20 ± 0.09 (8) | +0.07 ± 0.08 (5) |
| | EGG70 | 6210 | 4.5 | 1.36 | -0.07 | +0.00 ± 0.03 (1) | +0.10 ± 0.08 (3) | +0.16 ± 0.07 (5) | +0.25 ± 0.11 (7) | +0.00 ± 0.04 (5) |
| IC 4651 | AMC1109 | 6060 | 4.5 | 1.20 | 0.11 | +0.16 ± 0.03 (1) | +0.19 ± 0.05 (3) | +0.22 ± 0.05 (5) | +0.13 ± 0.06 (6) | +0.16 ± 0.03 (5) |
| | AMC2207 | 6050 | 4.4 | 1.18 | 0.13 | +0.13 ± 0.03 (1) | +0.17 ± 0.05 (2) | +0.16 ± 0.11 (5) | +0.16 ± 0.05 (4) | +0.11 ± 0.02 (5) |
| | AMC4220 | 5910 | 4.6 | 1.06 | 0.19 | +0.17 ± 0.04 (1) | +0.29 ± 0.11 (2) | +0.24 ± 0.09 (5) | +0.25 ± 0.02 (3) | +0.23 ± 0.07 (5) |

Table 6—Continued

| Cluster | Star | $T_{eff}$ | log $g$ | $\xi$ | [Fe/H] | $[Y/H]_{5087}$ | [Zr/H] | [La/H] | [Ce/H] | $[Y/H]_{all}$ |
|---|---|---|---|---|---|---|---|---|---|---|
| | AMC4226 | 5980 | 4.4 | 1.19 | 0.13 | +0.13 ± 0.04 (1) | +0.23 ± 0.06 (1) | +0.20 ± 0.11 (5) | +0.09 ± 0.04 (2) | +0.10 ± 0.04 (4) |
| | EGG45 | 6320 | 4.4 | 1.50 | 0.05 | +0.12 ± 0.03 (1) | +0.11 ± 0.05 (3) | +0.15 ± 0.09 (5) | +0.09 ± 0.09 (4) | +0.05 ± 0.06 (5) |
| M67 | S969 | 5800 | 4.4 | 1.10 | 0.01 | −0.05 ± 0.04 (1) | — | — | +0.03 ± 0.06 (2) | +0.00 ± 0.04 (5) |
| | S988 | 6153 | 4.1 | 1.45 | 0.03 | +0.03 ± 0.04 (1) | — | — | +0.08 ± 0.04 (2) | +0.04 ± 0.07 (5) |
| | S994 | 6151 | 4.1 | 1.45 | 0.00 | −0.02 ± 0.05 (1) | — | — | −0.05 ± 0.07 (2) | +0.03 ± 0.11 (5) |
| | S995 | 6210 | 3.9 | 1.50 | 0.05 | −0.03 ± 0.04 (1) | — | — | +0.04 ± 0.06 (2) | +0.04 ± 0.05 (4) |
| | S998 | 6223 | 4.0 | 1.50 | 0.07 | −0.07 ± 0.06 (1) | — | — | +0.07 ± 0.06 (1) | +0.00 ± 0.06 (4) |
| | S1034 | 6019 | 4.0 | 1.50 | 0.01 | +0.06 ± 0.04 (1) | — | — | +0.07 ± 0.08 (2) | +0.06 ± 0.08 (5) |
| | S1239 | 5541 | 3.8 | 1.25 | 0.02 | +0.01 ± 0.04 (1) | — | — | +0.06 ± 0.04 (2) | +0.06 ± 0.04 (4) |
| | S1252 | 5938 | 4.4 | 1.15 | 0.05 | +0.09 ± 0.05 (1) | — | — | +0.11 ± 0.07 (2) | +0.07 ± 0.04 (5) |
| | S1256 | 5907 | 4.5 | 1.10 | 0.06 | +0.03 ± 0.04 (1) | — | — | +0.03 ± 0.05 (2) | +0.04 ± 0.05 (5) |
| | S746 | 5750 | 4.4 | 0.84 | 0.07 | — | +0.09 ± 0.02 (3) | +0.06 ± 0.08 (5) | — | — |
| | S1048 | 5900 | 4.4 | 0.94 | 0.03 | — | — | +0.08 ± 0.08 (5) | +0.12 ± 0.06 (5) | — |
| | S1092 | 6160 | 4.4 | 1.42 | 0.07 | — | +0.05 ± 0.03 (3) | +0.15 ± 0.04 (5) | — | — |
| | S1255 | 5840 | 4.5 | 1.05 | 0.01 | — | +0.12 ± 0.04 (3) | +0.09 ± 0.04 (5) | — | +0.12 ± 0.06 (2) |
| | S1283 | 6100 | 4.4 | 1.16 | 0.03 | — | +0.06 ± 0.04 (3) | +0.08 ± 0.05 (5) | — | — |
| | S1287 | 6100 | 4.4 | 1.17 | -0.04 | — | +0.00 ± 0.07 (3) | +0.05 ± 0.04 (5) | — | — |
| NGC 6192 | 9 | 5230 | 2.8 | 1.95 | 0.21 | +0.36 ± 0.07 (1) | — | — | — | +0.52 ± 0.13 (4) |
| | 45 | 5050 | 2.6 | 1.65 | 0.04 | +0.19 ± 0.06 (1) | — | — | — | +0.37 ± 0.12 (5) |
| | 96 | 5050 | 2.5 | 2.10 | 0.08 | +0.13 ± 0.06 (1) | — | — | — | +0.47 ± 0.25 (4) |
| | 137 | 4670 | 1.9 | 1.80 | 0.00 | −0.01 ± 0.08 (1) | — | — | — | +0.37 ± 0.22 (5) |
| NGC 2324 | 10488 | 5100 | 2.2 | 1.21 | -0.17 | +0.06 ± 0.06 (1) | — | — | +0.01 ± 0.06 (2) | +0.18 ± 0.16 (4) |
| | 10787 | 5060 | 2.1 | 1.23 | -0.17 | +0.07 ± 0.07 (1) | — | — | +0.04 ± 0.08 (2) | +0.16 ± 0.11 (4) |
| | 10797 | 5040 | 2.2 | 1.21 | -0.17 | +0.09 ± 0.07 (1) | — | — | +0.06 ± 0.07 (2) | +0.23 ± 0.15 (5) |

Table 6—Continued

| Cluster | Star | $T_{\rm eff}$ | log $g$ | $\xi$ | [Fe/H] | [Y/H]$_{5087}$ | [Zr/H] | [La/H] | [Ce/H] | [Y/H]$_{\rm all}$ |
|---|---|---|---|---|---|---|---|---|---|---|
| | 10828 | 4750 | 1.6 | 1.29 | -0.10 | +0.08 ± 0.07 (1) | — | — | +0.06 ± 0.07 (2) | +0.37 ± 0.20 (5) |
| | 30878 | 5000 | 2.1 | 1.22 | -0.27 | −0.04 ± 0.08 (1) | — | — | −0.14 ± 0.08 (2) | +0.11 ± 0.13 (4) |
| | 30957 | 5000 | 1.8 | 1.26 | -0.17 | −0.15 ± 0.09 (1) | — | — | −0.18 ± 0.09 (2) | +0.16 ± 0.20 (5) |
| NGC 3960 | C4 | 5050 | 2.5 | 1.17 | 0.07 | +0.20 ± 0.08 (1) | — | — | +0.14 ± 0.05 (2) | +0.33 ± 0.13 (5) |
| | C5 | 4870 | 2.2 | 1.22 | 0.00 | +0.08 ± 0.07 (1) | — | — | +0.07 ± 0.06 (2) | +0.25 ± 0.13 (5) |
| | C6 | 4950 | 2.4 | 1.19 | 0.02 | +0.18 ± 0.07 (1) | — | — | +0.13 ± 0.05 (2) | +0.29 ± 0.13 (4) |
| | C8 | 5040 | 2.6 | 1.18 | 0.00 | +0.07 ± 0.07 (1) | — | — | +0.11 ± 0.05 (2) | +0.21 ± 0.15 (5) |
| | C9 | 5000 | 2.5 | 1.18 | 0.02 | +0.09 ± 0.08 (1) | — | — | +0.16 ± 0.06 (2) | +0.23 ± 0.11 (5) |
| NGC 2660 | 296 | 5200 | 3.0 | 1.11 | 0.08 | +0.31 ± 0.07 (1) | — | — | +0.41 ± 0.07 (2) | +0.43 ± 0.10 (4) |
| | 318 | 5030 | 2.6 | 1.16 | 0.00 | +0.01 ± 0.10 (1) | — | — | +0.23 ± 0.08 (2) | +0.33 ± 0.18 (5) |
| | 542 | 5060 | 2.5 | 1.17 | 0.03 | +0.05 ± 0.10 (1) | — | — | +0.14 ± 0.05 (2) | +0.31 ± 0.24 (5) |
| | 694 | 5100 | 2.8 | 1.14 | 0.05 | +0.28 ± 0.07 (1) | — | — | +0.37 ± 0.07 (2) | +0.41 ± 0.15 (4) |
| | 862 | 5100 | 2.6 | 1.16 | 0.02 | +0.30 ± 0.06 (1) | — | — | +0.25 ± 0.05 (2) | +0.31 ± 0.07 (5) |
| NGC 2477 | 36280 | 4980 | 2.8 | 1.14 | 0.05 | +0.32 ± 0.08 (1) | — | — | +0.23 ± 0.06 (2) | +0.39 ± 0.08 (5) |
| | 36288 | 4970 | 2.7 | 1.15 | 0.05 | +0.31 ± 0.08 (1) | — | — | +0.26 ± 0.06 (2) | +0.50 ± 0.12 (5) |
| | 36326 | 5000 | 2.7 | 1.15 | 0.10 | +0.37 ± 0.08 (1) | — | — | +0.26 ± 0.06 (2) | +0.51 ± 0.10 (5) |
| | 36363 | 4950 | 2.7 | 1.16 | 0.05 | +0.26 ± 0.06 (1) | — | — | +0.22 ± 0.05 (2) | +0.45 ± 0.12 (5) |
| | 36385 | 5030 | 2.7 | 1.15 | 0.07 | +0.10 ± 0.09 (1) | — | — | +0.09 ± 0.07 (2) | +0.44 ± 0.23 (5) |
| | 36449 | 4970 | 2.6 | 1.16 | 0.12 | +0.33 ± 0.08 (1) | — | — | +0.22 ± 0.06 (2) | +0.54 ± 0.13 (5) |
| Berkeley 29 | 257 | 4930 | 2.6 | 1.18 | -0.36 | −0.19 ± 0.09 (1) | — | — | — | −0.10 ± 0.06 (4) |
| | 398 | 5020 | 2.7 | 1.17 | -0.31 | +0.25 ± 0.08 (1) | — | — | — | +0.26 ± 0.08 (5) |
| | 602 | 4970 | 2.4 | 1.28 | -0.34 | −0.01 ± 0.09 (1) | — | — | −0.18 ± 0.09 (2) | +0.09 ± 0.22 (4) |
| | 933 | 4930 | 2.3 | 1.25 | -0.30 | +0.04 ± 0.09 (1) | — | — | −0.09 ± 0.09 (1) | +0.21 ± 0.10 (5) |
| Berkeley 20 | 1201 | 4700 | 2.1 | 1.35 | -0.28 | −0.53 ± 0.09 (1) | — | — | −0.43 ± 0.08 (2) | −0.66 ± 0.12 (5) |

Table 6—Continued

| Cluster | Star | $T_{\rm eff}$ | log $g$ | $\xi$ | [Fe/H] | [Y/H]$_{5087}$ | [Zr/H] | [La/H] | [Ce/H] | [Y/H]$_{\rm all}$ |
|---|---|---|---|---|---|---|---|---|---|---|
| | 1240 | 4400 | 1.7 | 1.27 | -0.31 | −0.32 ± 0.09 (1) | — | — | −0.19 ± 0.11 (2) | −0.06 ± 0.16 (5) |
| NGC 6253 | 23498 | 5630 | 3.4 | 1.30 | 0.32 | +0.52 ± 0.09 (1) | — | — | — | +0.60 ± 0.15 (5) |
| | 23501 | 6050 | 3.8 | 0.90 | 0.29 | +0.10 ± 0.10 (1) | — | — | — | +0.25 ± 0.30 (4) |
| | 24707 | 4940 | 3.0 | 1.16 | 0.39 | +0.11 ± 0.10 (1) | — | — | +0.35 ± 0.07 (2) | +0.48 ± 0.23 (5) |
| | 69885 | 6200 | 3.8 | 1.27 | 0.45 | +0.51 ± 0.09 (1) | — | — | — | +0.40 ± 0.16 (4) |
| | 105495 | 4450 | 2.5 | 1.23 | 0.49 | +0.28 ± 0.07 (1) | — | — | +0.38 ± 0.08 (2) | +0.67 ± 0.25 (5) |
| Melotte 66 | 1346 | 4750 | 2.0 | 1.17 | -0.37 | −0.30 ± 0.06 (1) | — | — | −0.41 ± 0.06 (2) | −0.22 ± 0.07 (5) |
| | 1493 | 4770 | 2.1 | 1.20 | -0.35 | −0.32 ± 0.06 (1) | — | — | −0.31 ± 0.05 (2) | −0.24 ± 0.08 (5) |
| | 1785 | 4770 | 2.0 | 1.20 | -0.30 | −0.40 ± 0.08 (1) | — | — | −0.40 ± 0.06 (2) | −0.25 ± 0.10 (5) |
| | 1865 | 4717 | 2.0 | 1.24 | -0.34 | — | — | — | −0.23 ± 0.06 (2) | +0.20 ± 0.09 (4) |
| | 1884 | 4750 | 2.5 | 1.23 | -0.30 | −0.20 ± 0.10 (1) | — | — | −0.16 ± 0.11 (2) | −0.08 ± 0.20 (5) |
| | 2212 | 4850 | 2.4 | 1.25 | -0.31 | −0.34 ± 0.07 (1) | — | — | −0.22 ± 0.07 (2) | −0.17 ± 0.12 (5) |
| Berkeley 32 | 17 | 4830 | 2.2 | 1.21 | -0.31 | −0.34 ± 0.07 (1) | — | — | −0.36 ± 0.06 (2) | −0.22 ± 0.09 (5) |
| | 18 | 4850 | 2.3 | 1.21 | -0.27 | −0.26 ± 0.07 (1) | — | — | −0.21 ± 0.07 (2) | −0.15 ± 0.06 (5) |
| | 19 | 4760 | 2.3 | 1.21 | -0.35 | −0.43 ± 0.08 (1) | — | — | −0.33 ± 0.07 (2) | −0.27 ± 0.12 (5) |
| | 25 | 4760 | 2.4 | 1.19 | -0.20 | −0.05 ± 0.09 (1) | — | — | −0.21 ± 0.07 (2) | −0.01 ± 0.08 (5) |
| | 27 | 4780 | 2.4 | 1.19 | -0.24 | −0.40 ± 0.09 (1) | — | — | −0.23 ± 0.09 (2) | −0.19 ± 0.19 (5) |
| | 45 | 4920 | 3.0 | 1.11 | -0.35 | −0.56 ± 0.10 (1) | — | — | — | −0.30 ± 0.15 (5) |
| | 938 | 4870 | 2.3 | 1.19 | -0.30 | −0.46 ± 0.09 (1) | — | — | −0.35 ± 0.06 (2) | −0.28 ± 0.12 (5) |
| | 940 | 4800 | 2.1 | 1.23 | -0.33 | −0.40 ± 0.09 (1) | — | — | −0.37 ± 0.10 (2) | −0.18 ± 0.13 (5) |
| | 941 | 4760 | 2.4 | 1.19 | -0.29 | −0.10 ± 0.07 (1) | — | — | −0.07 ± 0.06 (2) | +0.16 ± 0.17 (5) |
| Collinder 261 | 2 | 4350 | 1.7 | 1.25 | 0.12 | −0.11 ± 0.07 (1) | — | — | — | +0.53 ± 0.39 (5) |
| | 5 | 4600 | 2.0 | 1.24 | 0.14 | −0.18 ± 0.09 (1) | — | — | −0.08 ± 0.06 (2) | +0.14 ± 0.19 (5) |
| | 6 | 4500 | 2.3 | 1.18 | 0.16 | +0.01 ± 0.07 (1) | — | — | — | +0.43 ± 0.26 (5) |

# Table 6—Continued

| Cluster | Star | $T_{\rm eff}$ | log $g$ | $\xi$ | [Fe/H] | $[Y/H]_{5087}$ | [Zr/H] | [La/H] | [Ce/H] | $[Y/H]_{\rm all}$ |
|---|---|---|---|---|---|---|---|---|---|---|
| | 7 | 4546 | 2.1 | 1.20 | 0.18 | −0.03 ± 0.09 (1) | — | — | — | +0.43 ± 0.32 (5) |
| | 9 | 4720 | 2.0 | 1.27 | 0.04 | −0.09 ± 0.07 (1) | — | — | −0.09 ± 0.07 (2) | +0.05 ± 0.14 (5) |
| | 10 | 4700 | 2.4 | 1.20 | 0.20 | +0.06 ± 0.09 (1) | — | — | — | +0.32 ± 0.23 (5) |
| | 11 | 4670 | 2.1 | 1.13 | 0.09 | −0.17 ± 0.09 (1) | — | — | −0.05 ± 0.08 (2) | +0.15 ± 0.19 (5) |

Note. — Column 7 shows yttrium abundances from the 508.742 nm line only. Column 11 shows yttrium abundances derived through all the available lines

Table 7. Comparison of stellar parameters

| Star name | | Photometry-literature | Spectroscopy-literature | Spectroscopy-this study |
|---|---|---|---|---|
| KW208 | $T_{eff}$ | 5993 | 6208 | 6150 |
| | $\log g$ | 4.45 | 4.58 | 4.44 |
| | $\xi$ | — | 1.52 | 1.43 |
| | ♯ FeI | — | — | 31 |
| | ♯ FeII | — | — | 6 |
| | [Y/H] | — | 0.28 ± 0.05 | 0.20 ± 0.05 |
| KW392 | $T_{eff}$ | 5902 | 6250 | 6070 |
| | $\log g$ | 4.46 | 4.56 | 4.28 |
| | $\xi$ | | 1.48 | 1.38 |
| | ♯ FeI | — | — | 36 |
| | ♯ FeII | — | — | 8 |
| | [Y/H] | — | 0.31 ± 0.06 | 0.21 ± 0.05 |
| TATM11003 | $T_{eff}$ | 5625 | 6160 | 6210 |
| | $\log g$ | | 4.74 | 4.59 |
| | $\xi$ | | 1.05 | 1.08 |
| | ♯ FeI | — | — | 30 |
| | ♯ FeII | — | — | 7 |
| | [Ce/H] | — | 0.27 ± 0.06 | 0.24 ± 0.05 |
| TATM11014 | $T_{eff}$ | 5764 | 6273 | 6120 |
| | $\log g$ | | 4.74 | 4.39 |
| | $\xi$ | | 1.26 | 1.34 |
| | ♯ FeI | — | — | 28 |

Table 7—Continued

| Star name | Photometry-literature | Spectroscopy-literature | Spectroscopy-this study |
|---|---|---|---|
| ♯ FeII | — | — | 4 |
| [Ce/H] | — | 0.32 ± 0.04 | 0.22 ± 0.04 |

Note. — Literature data are from Pace et al. (2008) for KW 208 and KW 392 and from Pace et al. (2010) and references therein for TATM 11003 and TATM 11014.